\documentclass[12pt]{article}
\usepackage{amsfonts,amsthm,amsmath,amssymb,upgreek,bm}
\usepackage{float}
\usepackage{cite}
\usepackage[colorlinks=true, linkcolor=black, citecolor=cyan, urlcolor=cyan]{hyperref}
\usepackage[paper=letterpaper,margin=.85in]{geometry}
\usepackage{graphicx}
\usepackage{units}
\usepackage[export]{adjustbox}
\usepackage{xcolor}
\usepackage[shortlabels]{enumitem}
\usepackage[mathscr]{euscript}
\usepackage[normalem]{ulem}
\usepackage{tikz}
\usepackage{tikzsymbols}
\usepackage{pgfplots}
\pgfplotsset{compat=1.18}
\usepackage{ytableau}
\usepackage{slashed}
\usepackage{mathtools}
\usepackage{bbold}
\usepackage{comment}

\usetikzlibrary{decorations.pathmorphing,arrows.meta,calc}

\newcommand{\K}{\mathbb K}
\newcommand{\Z}{\mathbb Z}
\newcommand{\ii}{\mathrm{i}}
\newcommand{\N}{\mathcal{N}}
\DeclareMathOperator*{\Res}{Res}
\numberwithin{equation}{section}

\def\s{\mathfrak{s}}

\definecolor{teall}{HTML}{0FBFB3}

\begin{document}
\thispagestyle{empty}
\vspace*{2.5cm}
\begin{center}
{\bfseries\LARGE Cusp regularization with spin structures \\
\vspace{.5cm}
and gravitational anomalies}
\begin{center}
\vspace{1cm}
Elisa Tabor$^{1}$ and Cynthia Yan$^{2}$\\
\bigskip
\bigskip
$^{1}$Leinweber Institute for Theoretical Physics,\\
Stanford University, Stanford, CA 94305\\
\bigskip
$^{2}$Jefferson Physical Laboratory,\\
Harvard University, Cambridge, MA 02138
\end{center}
\vspace{2.5cm}
{\bfseries Abstract}
\end{center}
\begin{quotation}
\noindent

The family of $SL(2,\mathbb Z)$ black holes has an accumulation point corresponding to a cusped geometry, which causes the sum over black holes to diverge. We extend cusp regularization to the sum over geometries compatible with a fixed boundary spin structure. For $c_L=c_R$, including the cusp as a counterterm cancels the leading divergence for each of the three even boundary spin structures. The required subtraction is the same in all three sectors and is two thirds of the bosonic one. We further show that gravitational anomalies cause cancellations between black hole contributions, removing the leading cusp divergence except at $c_L-c_R=0,\pm24$ in pure gravity and $c_L-c_R=0,\pm12$ in $\mathcal N=(1,1)$ supergravity. We also derive the spin structure dependent anomaly phases responsible for these cancellations.
\end{quotation}
\setcounter{page}{0}
\setcounter{tocdepth}{2}
\setcounter{footnote}{0}
\newpage
\parskip 0.1in
\setcounter{page}{2}
\tableofcontents
\newpage

\section{Introduction}
Pure three-dimensional gravity has no local graviton degrees
of freedom. In anti-de Sitter (AdS) space, it nevertheless
has boundary excitations and black holes
\cite{Brown:1986nw,Banados:1992wn}. Defining their quantum
theory and its two-dimensional conformal field theory (CFT)
description is a longstanding problem \cite{Witten:2007kt}.
A gravitational path-integral description requires amplitudes
for individual geometries and a prescription for summing
their contributions.

The matrix-integral description of Jackiw-Teitelboim (JT)
gravity \cite{Saad:2019lba} has motivated statistical approaches
to three-dimensional gravity. These approaches build on
universal CFT thermodynamics \cite{Hartman:2014oaa} and the
averaged behavior of heavy-operator data
\cite{Kraus:2016nwo,Cardy:2017qhl,Das:2017cnv,Collier:2019weq,Belin:2021ryy,Anous:2021caj,Belin:2021ibv}.\footnote{These
descriptions specify statistical properties of CFT data,
not a microscopic ensemble of exact CFTs.}
This viewpoint relates on-shell wormholes to correlations
between CFT observables
\cite{Belin:2020hea,Chandra:2022bqq,Yan:2023rjh,deBoer:2024mqg,abdalla:wip},
while off-shell torus wormholes probe energy-level correlations
\cite{Cotler:2020ugk,Cotler:2020hgz,DiUbaldo:2023qli,Boruch:2025ilr}.
Frameworks for gravitational amplitudes and sums include
Virasoro topological quantum field theory
\cite{Collier:2023fwi,Collier:2024mgv,Belin:2026pko}, conformal Turaev-Viro
theory \cite{Hartman:2025cyj,Hartman:2025ula}, and
matrix-tensor models \cite{Belin:2023efa,Jafferis:2025vyp, Jafferis:2026bxj}.

Here we revisit the older problem of regularizing the Maloney-Witten sum over $SL(2,\mathbb Z)$ black holes with a fixed torus boundary \cite{Maloney:2007ud}. These are thermal AdS and its modular images, distinguished by which boundary cycle contracts in the bulk \cite{Maldacena:1998bw,Dijkgraaf:2000fq}. Their perturbative contributions are known \cite{Giombi:2008vd}, but the infinite sum diverges. Analytic continuation defines it by a form of zeta-function regularization \cite{Maloney:2007ud,Keller:2014xba}. Stanford-Yan gave this divergence a geometric interpretation, showing that the smooth fillings accumulate toward a noncompact cusped geometry \cite{Stanford:2025llj}. They regulated this accumulation by smoothly suppressing fillings with increasingly long cusp-like regions and including the limiting cusp as a counterterm, an additional contribution that cancels the divergence. This gives a geometric interpretation of zeta-function regularization.

Our first main result is that we extend this cusp prescription to gravity with a specified boundary spin structure, as considered by Boruch-Tabor-Turiaci (BTT) \cite{Boruch:2026hbr}. A spin structure specifies periodic or antiperiodic boundary conditions for a hypothetical test fermion around each torus cycle. The corresponding torus partition functions transform into one another under modular transformations \cite{Benjamin:2020zbs}. For a spin structure to extend over a bulk geometry, a cycle that contracts to a disk must be antiperiodic. Fixing the boundary spin structure therefore restricts the black hole sum. Does a cusp carrying that spin structure still cancel the divergence? This is a question about gravity with spin structures, not supergravity: we introduce neither dynamical fermion matter nor a gravitino for this first result.\footnote{For JT gravity with spin structures and its late-time correlators, see \cite{Stanford:2019vob,Yan:2022nod}.}

We show that the cusp cancels the leading divergence for each of the three even boundary spin structures, which are the spin structures for which at least one torus cycle can contract in a smooth bulk filling. To compare with the bosonic prescription, write the modulus as $\tau=\tau_1+i\tau_2$, with $\tau_2>0$, and take $c_L=c_R=c$. Let $q=e^{2\pi\ii\tau}$, let $\eta$ be the Dedekind eta function, and let $\epsilon>0$ be the cutoff parameter. The pair $(r,d)$ labels the geometry in which $r\tau+d$ contracts; we sum over relatively prime pairs, counting $(r,d)\sim(-r,-d)$ once. The symbol $\big|_\gamma$ evaluates the whole summand at the associated modular image $\gamma\tau$.
The bosonic cutoff sum and cusp subtraction are
\begin{equation}
Z_{\rm bos}(\tau) =\frac{1}{\sqrt{\tau_2}\,|\eta(\tau)|^2}
\left[
-\frac{6}{\sqrt{\pi\epsilon}}
+\sum_{(r,d)=1}
\left(
\sqrt{\tau_2}\,|q|^{-(c-1)/12}|1-q|^2
e^{-\epsilon/\tau_2}
\right)\bigg|_\gamma
\right].
\label{eq:intro-bosonic}
\end{equation}
For any of the three even spin structures, we obtain instead
\begin{equation}
Z_{\s}(\tau) = \frac{1}{\sqrt{\tau_2}\,|\eta(\tau)|^2}
\left[
-\frac{4}{\sqrt{\pi\epsilon}}
+\sum_{\substack{(r,d)=1\\\text{spin-compatible}}}
\left(
\sqrt{\tau_2}\,|q|^{-(c-1)/12}|1-q|^2
e^{-\epsilon/\tau_2}
\right)\bigg|_\gamma
\right].
\label{eq:main-result}
\end{equation}
In both expressions, the first term in the brackets is the cusp subtraction, and the second is the regulated black hole sum. Compared with \eqref{eq:intro-bosonic}, the sum is restricted to cycles compatible with the chosen spin structure, and six is replaced by four. We will see that the factor of two thirds arises because each even spin structure allows precisely two of the three parity classes of $(r,d)$, which contribute equally to the leading divergence. The local graviton contribution and the cusp determinant are unchanged, and the subtraction coefficient is the same for all three choices of even spin structure. We determine it by using a Mellin transform to relate the cutoff divergence to the pole of the parity-restricted sum.

Our second main result is that a gravitational anomaly generally removes the leading cusp divergence. Such anomalies can be incorporated in the bulk through invertible topological theories \cite{Witten:2015aba}, whose allowed phases are constrained by cobordism \cite{Kapustin:2014dxa} and can depend on the spin structure \cite{Stanford:2019vob, Turiaci:2023jfa, Boruch:2026hbr}. When the left- and right-moving central charges differ, the gravitational anomaly \cite{Kraus:2005zm} introduces relative phases between black hole fillings \cite{Witten:2015aba, Boruch:2026hbr}, preventing their contributions from adding coherently. We find that the leading divergence survives only for $c_L-c_R=0,\pm24$ in pure gravity and $c_L-c_R=0,\pm12$ in $\mathcal N=(1,1)$ supergravity. These cases arise when the anomalous angular phase cancels that of a vacuum-character term. One important ingredient is the fermionic gravitational anomaly phase, derived in Eq.~\eqref{eq:spin-anomaly-phase} in terms of modified Rademacher functions.

In the bosonic case, taking $\epsilon\to0$ in \eqref{eq:intro-bosonic} reproduces the zeta-regularized Maloney-Witten answer. For the spin structure restricted and gravitationally anomalous sums, the results established here are the determination of the leading divergences and their cancellation when present. This should be distinguished from the separate problem of negative spectral weights in the finite answer \cite{Keller:2014xba,Benjamin:2020mfz}.\footnote{Proposals to address spectral negativity include conical defects \cite{Benjamin:2020mfz} and additional off-shell topologies \cite{Maxfield:2020ale}. See also \cite{Benjamin:2019stq, DiUbaldo:2023hkc, Chang:2026ylb}.}

The paper is organized as follows. Section~\ref{sec:spin} introduces the boundary spin structures and their allowed black hole geometries. Section~\ref{sec:tau} derives the cusp subtraction for each even spin structure. Section~\ref{sec:unequal-central-charges} determines the effect of gravitational anomalies on the cusp divergence in pure gravity and in $\mathcal N=(1,1)$ supergravity. Appendix~\ref{app:parity} supplies the detailed proofs for the parity-restricted sums, while Appendix~\ref{app:energy} checks the subtraction in the energy basis and distinguishes it from the finite negative threshold weights found by BTT. Appendix~\ref{app:anomalous-pole} proves the Eisenstein pole selection rule in the presence of gravitational anomalies, and Appendix~\ref{app:spin-anomaly} explains the bulk origin of the spin structure anomaly phase in terms of gravitational Chern-Simons theory and spin lens-space invariants.

\section{Spin structures and the SL\texorpdfstring{$(2,\mathbb Z)$}{(2,Z)} black holes}
\label{sec:spin}

Before regularizing the black hole sum, we must specify the allowed solid-torus fillings of the boundary torus, distinguished by which boundary cycle contracts in the bulk. Fixing a spin structure restricts the allowed fillings, since the boundary spin structure must extend smoothly into the bulk. We review BTZ and its modular images, then determine which fillings are compatible with each boundary spin structure. 

\begin{figure}[H]
\centering
\newcommand{\blackholetorusshape}{%
    (211.035,258.055)
    .. controls (312.101,258.055) and (394.154,207.441) .. (394.154,145.099)
    .. controls (394.154,82.7575) and (312.101,32.1438) .. (211.035,32.1438)
    .. controls (109.97,32.1438) and (27.9169,82.7575) .. (27.9169,145.099)
    .. controls (27.9169,207.441) and (109.97,258.055) .. (211.035,258.055)
    -- cycle
    (134.238,146.305)
    .. controls (158.606,130.386) and (214.118,104.461) .. (274.362,146.302)
    .. controls (199.735,202.928) and (134.238,146.305) .. (134.238,146.305)
    -- cycle
}
\newcommand{\blackholetorusbody}{%
  \path[fill={rgb,1:red,.5;green,.5;blue,.5},
        fill opacity=.9,even odd rule] \blackholetorusshape;
  \draw[black]
    (211.035,258.055)
    .. controls (312.101,258.055) and (394.154,207.441) .. (394.154,145.099)
    .. controls (394.154,82.7575) and (312.101,32.1438) .. (211.035,32.1438)
    .. controls (109.97,32.1438) and (27.9169,82.7575) .. (27.9169,145.099)
    .. controls (27.9169,207.441) and (109.97,258.055) .. (211.035,258.055)
    -- cycle;
  \draw[black] (121.221,156.015)
    .. controls (121.221,156.015) and (201.954,82.4948) .. (288.163,157.046);
  \draw[black] (134.238,146.305)
    .. controls (134.238,146.305) and (199.735,202.928) .. (274.362,146.302);
}
\pgfmathdeclarefunction{torusrimfit}{3}{%
  \pgfmathparse{exp(-((360*sin((#1-#2)/2)/(pi*#3))^2))}%
}
\newcommand{\blackholewinding}[3][]{%
  \draw[red,#1] plot[smooth,samples=241,variable=\t,domain=#2:#3]
    ({211.035+(126.5+56.5*cos(2*\t))*cos(\t+45)
       -11.5*torusrimfit(\t,105.383087,14)-.9*torusrimfit(\t,24,16)
       +3.5*torusrimfit(\t,164.6,13)},
     {145.099+.526*(126.5+56.5*cos(2*\t))*sin(\t+45)
       +35.8*sin(2*\t)-.744443*torusrimfit(\t,105.383087,14)
       -2.3*torusrimfit(\t,243,20)-4*torusrimfit(\t,24,16)
       +3.8*torusrimfit(\t,164.6,13)});
}
\newcommand{\blackholecyclepanel}[1]{%
  \begin{tikzpicture}[x=1bp,y=1bp,line width=3bp,
                      line cap=round,line join=round]
    \path[use as bounding box] (0,0) rectangle (419.76,297.84);
    \ifnum#1=0
      \draw[red] (134.238,146.305)
        arc[start angle=60.027271,end angle=-119.972729,
            radius=46.347514bp];
    \fi
    \blackholetorusbody
    \ifnum#1=0
      \draw[red] (87.9287,66.0067)
        arc[start angle=240.027271,end angle=60.027271,
            radius=46.347514bp];
    \else
      \blackholewinding[draw={rgb,1:red,.55;green,.45;blue,.45}]{0}{360}
      \blackholewinding{24.196923}{105.383087}
      \blackholewinding{164.616913}{245.803077}
    \fi
  \end{tikzpicture}%
}
\begin{minipage}[b]{.43\textwidth}
\centering
\resizebox{!}{3.7cm}{\blackholecyclepanel{0}}
\par\small (a) BTZ
\end{minipage}
\hspace{.04\textwidth}
\begin{minipage}[b]{.43\textwidth}
\centering
\resizebox{!}{3.7cm}{\blackholecyclepanel{1}}
\par\small (b) $(r,d)=(2,1)$ black hole
\end{minipage}
\caption{Red marks the boundary cycle chosen to contract in each
filling: (a) $\tau$; (b) $2\tau+1$.}
\label{fig:black-hole-fillings}
\end{figure}
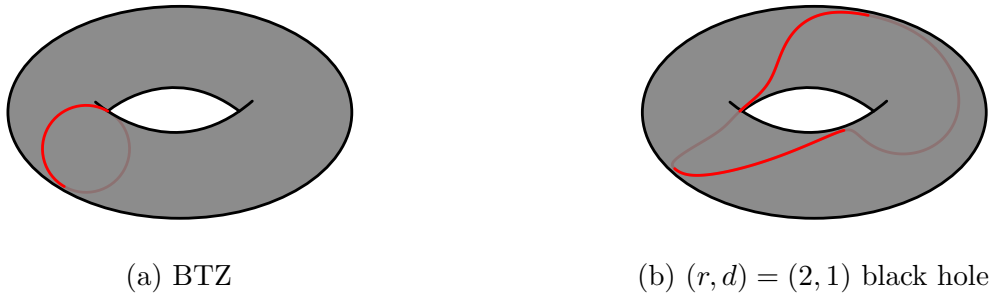

\paragraph{BTZ and its modular images.}
Fix a torus boundary of modulus $\tau$, with spatial cycle
$1$ and Euclidean-time cycle $\tau$. The Euclidean BTZ black
hole is a solid torus in which the time cycle contracts,
while the spatial circle remains noncontractible. In thermal
AdS, the spatial cycle contracts instead. More generally,
the $SL(2,\mathbb Z)$ black holes correspond to different
choices of a primitive contractible cycle
\cite{Maloney:2007ud,Yan:2025usw}:
\begin{equation}
\gamma=\begin{pmatrix}a&b\\r&d\end{pmatrix}\in SL(2,\mathbb Z),
\qquad
\gamma\tau=\frac{a\tau+b}{r\tau+d}.
\label{eq:gamma}
\end{equation}
The cycle $r\tau+d$ winds $r$ times around time and $d$ times
around space, with $(r,d)=1$. Opposite orientations describe
the same filling, so $(r,d)\sim(-r,-d)$; we take $r>0$,
together with $(r,d)=(0,1)$. BTZ has $(r,d)=(1,0)$ and thermal
AdS has $(r,d)=(0,1)$. Figure~\ref{fig:black-hole-fillings}
shows how the same boundary spatial circle sits in different
solid-torus fillings.

\paragraph{The boundary spin structure: what do the labels mean?}
Following BTT \cite{Boruch:2026hbr}, we consider pure
gravity with a spin structure, without dynamical fermion
matter. We describe the spin structure by asking how a
hypothetical test fermion would behave around each boundary
cycle. This probe only describes the geometric data; it adds
no contribution to the partition function.

A spin structure specifies whether the test fermion would be
periodic or antiperiodic around each cycle. The spin structure can be labeled by
$(\mu,\nu)$ where $\mu$ refers to the time cycle and $\nu$ to the
spatial cycle, with $0$ meaning periodic and $1$ antiperiodic. Here we will refer to spin structures using $R^\pm$ and $NS^\pm$, which encode the same information. The letter refers to \emph{space}: Ramond ($R$) means periodic
boundary conditions, so $\nu=0$, while Neveu-Schwarz ($NS$)
means antiperiodic boundary conditions, so $\nu=1$.
The superscript refers to
\emph{time}: $+$ means periodic, so $\mu=0$, and $-$ means
antiperiodic, so $\mu=1$. In the boundary Hilbert-space
description, $-$ gives the ordinary thermal trace, whereas
$+$ inserts fermion parity $(-1)^{\sf F}$ in the trace,
counting bosonic states minus fermionic states.

\begin{figure}[H]
\centering
\begin{tikzpicture}[
  font=\small,
  can contract/.style={draw=blue!55!black,line width=1.3pt,
    -{Stealth[length=4pt,width=4pt]}},
  cannot contract/.style={draw=black!45,line width=.8pt,dashed,
    -{Stealth[length=4pt,width=4pt]}}
]
\foreach \panelx/\sector/\spin/\tauperiodicity/\oneperiodicity/\taustyle/\onestyle/\diagonalstyle in {
  0/{R^-}/{1,0}/antiperiodic/periodic/can contract/cannot contract/can contract,
  3.95/{NS^+}/{0,1}/periodic/antiperiodic/cannot contract/can contract/can contract,
  7.9/{NS^-}/{1,1}/antiperiodic/antiperiodic/can contract/can contract/cannot contract,
  11.85/{R^+}/{0,0}/periodic/periodic/cannot contract/cannot contract/cannot contract%
}{
  \begin{scope}[xshift=\panelx cm]
    \node[font=\normalsize] at (1.175,3.18) {$\sector$};
    \node at (1.175,2.76) {$(\mu,\nu)=(\spin)$};
    \filldraw[fill=black!2,draw=black!25,line width=.5pt]
      (0,0) rectangle (2.35,2.35);
    \draw[\onestyle] (0,0) -- (2.35,0);
    \draw[\taustyle] (0,0) -- (0,2.35);
    \draw[\diagonalstyle] (0,0) -- (2.35,2.35)
      node[pos=.58,above,sloped,fill=white,inner sep=1.5pt]
      {$\tau+1$};
    \node[rotate=90] at (-.36,1.175)
      {$\tau$: \tauperiodicity};
    \node at (1.175,-.35) {$1$: \oneperiodicity};
  \end{scope}
}
\draw[can contract] (0,-1.02) -- (.65,-1.02)
  node[right,black] {can contract (antiperiodic)};
\draw[cannot contract] (7.6,-1.02) -- (8.25,-1.02)
  node[right,black] {cannot contract (periodic)};
\end{tikzpicture}
\caption{Four boundary spin structures, shown as squares with
opposite edges identified.
Solid blue (antiperiodic) cycles can contract; dashed gray
(periodic) cycles cannot.}
\label{fig:spin-structures}
\end{figure}
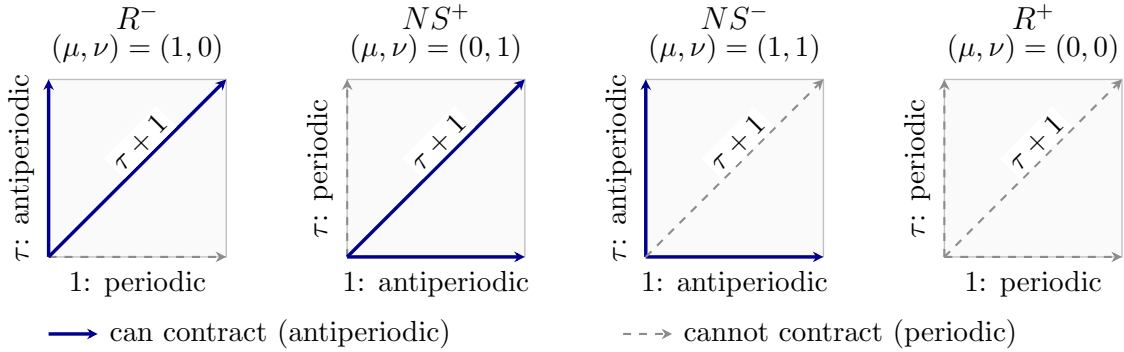

\paragraph{Which black hole geometries are compatible?}
We now hold this boundary spin structure fixed and ask which
of the cycles $r\tau+d$ can contract in the bulk.
Going around $r\tau+d$ would multiply the test fermion by
$(-1)^{r\mu}(-1)^{d\nu}=(-1)^{r\mu+d\nu}$.
For the spin structure to extend smoothly over the disk
bounded by this cycle, the test fermion must have antiperiodic
boundary conditions. The filling is therefore allowed precisely when
\begin{equation}
r\mu+d\nu=1\pmod2.
\label{eq:bounding}
\end{equation}
For $R^-$, only a winding around the time cycle contributes a
minus sign, so $r$ must be odd. For $NS^+$, only a spatial
winding contributes a minus sign, so $d$ must be odd. For
$NS^-$, both do, so $r+d$ must be odd. For $R^+$, no winding
can give a minus sign, and no solid-torus filling is allowed.
Figure~\ref{fig:spin-structures} illustrates which cycles can
contract for each boundary spin structure.
To summarize, $(\mu,\nu)$ specifies the \emph{boundary
conditions}, while $r$ and $d$ count the windings of the
contractible cycle. Equation~\eqref{eq:bounding} gives the
following restrictions on the primitive winding numbers:
\begin{equation}
\begin{array}{c|c|c|l}
\text{sector}&(\mu,\nu)&\text{(time, space)}&\text{condition}\\ \hline
R^-&(1,0)&\text{(can contract, non-contractible)}&r\ \text{odd}\\
NS^+&(0,1)&\text{(non-contractible, can contract)}&d\ \text{odd}\\
NS^-&(1,1)&\text{(can contract, can contract)}&r+d\ \text{odd}\\
R^+&(0,0)&\text{(non-contractible, non-contractible)}&\text{no solid-torus filling}
\end{array}
\label{eq:sectors}
\end{equation}
The three choices $R^-,NS^+,NS^-$ are the even spin
structures, and the fourth choice, $R^+$, is the odd spin structure. Since a cycle with $r$ and $d$ both even cannot be primitive, there are three primitive parity classes. Each even spin structure allows precisely two of these primitive parity classes. For example, thermal AdS, with $(r,d)=(0,1)$, contributes to both NS sectors but neither R sector. The usual BTZ filling, with $(r,d)=(1,0)$, contributes only to $R^-$ and $NS^-$. 

The spin structure restricts the allowed black hole geometries, while each allowed geometry contributes the usual graviton amplitude. In particular, for a fixed even spin structure $\s$, the sum over compatible solid-torus fillings takes the form \cite{Boruch:2026hbr}
\begin{equation} \label{eq:Z-spin}
    Z_{\s}(\tau) = \sum_{\gamma \in \mathbb{Z} \backslash \Gamma_{\s}} \, \chi^L_{\s}(\gamma \tau)\bar{\chi}^R_{\s}(\gamma \bar{\tau})\,,
\end{equation}
where $\Gamma_{\s}$ is the subgroup of $SL(2,\Z)$ that preserves the chosen spin structure, leaving its periodic or antiperiodic boundary conditions around both cycles unchanged \cite{Maloney:2007ud}. It is given by
\begin{align} \label{eq:spin-Gamma}
    \Gamma_{R-} &= \Gamma_0(2) \equiv \{ g \in \text{PSL}(2,\Z)\, | \, r = 0 \text{ mod }2\}\nonumber \\
    \Gamma_{NS+} &= \Gamma^0(2) \equiv \{ g\in \text{PSL}(2,\Z) \, | \, b=0 \text{ mod }2\} \\
    \Gamma_{NS-} &= \Gamma_ \theta \equiv \{ g\in\text{PSL}(2,\Z) \, | \, a+b = r + d = 1 \text{ mod }2\}\,. \nonumber
\end{align}
In the case of pure gravity, the summand in~\eqref{eq:Z-spin} uses $\chi_{\s}(\tau)=\chi_{\rm vac}(\tau)$ for $NS^\pm$, where $\chi_{\rm vac}(\tau)$ is the same as the bosonic vacuum character. For $R^-$, $\chi_{\s}(\tau)=\chi_{\rm vac}(S\tau)$ is a function rather than a character.

\section{Cusp regularization in the \texorpdfstring{$\tau$}{tau} basis}
\label{sec:tau}

The torus partition function involves an infinite sum over
the black hole geometries.
Selecting the allowed geometries does not make this sum
convergent. In the bosonic theory, Maloney-Witten define the
sum by analytic continuation \cite{Maloney:2007ud}.
Stanford-Yan give this regularization a geometric
interpretation: the geometries approach an accumulation
point described by a cusp, whose contribution cancels the
divergence of a smoothly cut-off sum \cite{Stanford:2025llj}.
Here we ask how this cusp subtraction changes when a
boundary spin structure is fixed.

The spin restrictions remove some geometries but leave the
same cusp accumulation point. No cycle contracts in the
limiting cusp, so each boundary spin structure extends
over it. In the pure Einstein theory considered here, the
one-loop determinant involves metric fluctuations, not
spinor fields, and is insensitive to spin structure.
We can therefore use the same cusp determinant.\footnote{As in \cite{Stanford:2025llj}, interpreting this one-loop result as the
complete cusp contribution assumes that higher loops vanish
or only change its overall normalization.}
The full cusp contribution for an even boundary spin
structure therefore has the form
\begin{equation}
Z_{\rm cusp}^{\text{even spin}}(\tau)\propto
\frac1{\sqrt{\tau_2}\,|\eta(\tau)|^2}
\label{eq:cusp}
\end{equation}
where $\eta(\tau)$ is the Dedekind eta function. The cusp contribution comes with an a priori undetermined fugacity, which we fix by requiring it to cancel the leading divergence of the spin-restricted sum.

The computation proceeds in three steps.
Section~\ref{sec:bosonic-regularization} reviews zeta-function
and cusp regularization of the bosonic black hole sum.
Section~\ref{sec:mellin-divergence} shows how a Mellin transform
converts the pole of the sum used in zeta-function
regularization into the cutoff divergence, fixing the
bosonic cusp subtraction.
Section~\ref{sec:spin-counterterm} then imposes the spin structure
restrictions, computes the divergence from each allowed
parity class, and fixes the cusp counterterm. Its coefficient
is the same for all three even spin structures, but differs
from the bosonic one.

\subsection{Review of the black hole sum and its regularization}
\label{sec:bosonic-regularization}

We first recall the unrestricted bosonic sum studied by
Maloney-Witten and Stanford-Yan
\cite{Maloney:2007ud,Stanford:2025llj}.
Including the vacuum and its boundary-graviton excitations
\cite{Giombi:2008vd},
the formal sum over black hole geometries is
\begin{equation}
Z(\tau)=\frac1{\sqrt{\tau_2}\,|\eta(\tau)|^2}
\sum_{(r,d)=1}
\left(\sqrt{\tau_2}\,
|q|^{-(c-1)/12}|1-q|^2\right)|_\gamma.
\label{eq:vac-cusp}
\end{equation}
Here $q=e^{2\pi \ii\tau}$, and the sum counts each primitive
pair $(r,d)\sim(-r,-d)$ once. The vertical bar means evaluation at
the transformed modulus:
\begin{equation}
f(\tau,\bar\tau)|_\gamma=f(\gamma\tau,\gamma\bar\tau),
\qquad
\tau_2|_\gamma=\frac{\tau_2}{|r\tau+d|^2}.
\label{eq:slash}
\end{equation}
where $\tau=\tau_1+\ii\tau_2$. In particular, all the $q$'s inside the parentheses in~\eqref{eq:vac-cusp} are
evaluated at $\gamma\tau$. This form follows from the modular
transformation of $\eta$: the combination
$\sqrt{\tau_2}|\eta(\tau)|^2$ is modular invariant.

The sum~\eqref{eq:vac-cusp} diverges; a prescription is
needed to define $Z(\tau)$. Maloney-Witten use zeta-function regularization:
\begin{equation}
Z(\tau)=\frac1{\sqrt{\tau_2}\,|\eta(\tau)|^2}
\mathop{\mathrm{continuation}}_{s\to1/2}
\sum_{(r,d)=1}
\left(\tau_2^s\,|q|^{-(c-1)/12}|1-q|^2\right)|_\gamma.
\label{eq:bosonic-zeta}
\end{equation}
The only change in the summand is
$\sqrt{\tau_2}\to\tau_2^s$. The modified sum converges for
$\Re s>1$ and is then analytically continued to
$s=\tfrac12$.

Stanford-Yan write the same answer using cusp regularization:
a smooth cutoff together with a cusp contribution,
\begin{equation}
Z(\tau)=\frac1{\sqrt{\tau_2}\,|\eta(\tau)|^2}
\lim_{\epsilon\to0}
\left[
-\frac6{\sqrt{\pi\epsilon}}
+\sum_{(r,d)=1}
\left(\sqrt{\tau_2}\,
|q|^{-(c-1)/12}|1-q|^2e^{-\epsilon/\tau_2}\right)|_\gamma
\right].
\label{eq:bosonic-cutoff}
\end{equation}
For $\epsilon>0$, the exponential becomes
$e^{-\epsilon|r\tau+d|^2/\tau_2}$ and suppresses long
cycles. The cusp term cancels the divergence when the cutoff
is removed.
We next review how the coefficient $-6/\sqrt{\pi\epsilon}$ follows from the sum,
before imposing the spin restrictions.

\subsection{Relating zeta-function and cusp regularization}
\label{sec:mellin-divergence}

We now derive the bosonic subtraction coefficient
$-6/\sqrt{\pi\epsilon}$, following Appendix B of Stanford-Yan. To compare the two regularizations, we first isolate the divergent part of the sum.

Expand $|1-q|^2=1+q\bar q-q-\bar q$. For long cycles,
$\tau_2|_\gamma\to0$, and the two scalar terms satisfy
\[
\left(|q|^{-(c-1)/12}(1+q\bar q)\right)|_\gamma
=2+O(\tau_2|_\gamma).
\]
The correction gives an absolutely convergent sum, while
the terms containing $-q-\bar q$ converge through phase
cancellations. These pieces have the same finite limit in
both prescriptions. Only the constant $2$ needs a subtraction.

After setting aside this factor of $2$ and the common
prefactor $1/(\sqrt{\tau_2}|\eta(\tau)|^2)$, the two sums are
\begin{equation} \label{eq:EF}
\begin{split}
E(\tau,s)&=\sum_{(r,d)=1}\tau_2^s|_\gamma,\\
F(\tau,\epsilon)&=\sum_{(r,d)=1}
\left(\sqrt{\tau_2}\,e^{-\epsilon/\tau_2}\right)|_\gamma.
\end{split}
\end{equation}
Here $E(\tau,s)$ is the sum used in zeta-function regularization,
while $F(\tau,\epsilon)$ is the cutoff sum used in cusp regularization. Zeta-function regularization assigns a finite value to the divergent sum by analytically continuing $E(\tau,s)$ to $s=\frac12$; cusp regularization instead subtracts the divergence of $F(\tau,\epsilon)$ before taking $\epsilon\to0$.

Mellin inversion expresses the exponential cutoff in terms
of the powers used in $E(\tau,s)$. Applying it to each term of $F(\tau,\epsilon)$
and summing gives
\begin{equation}
F(\tau,\epsilon)
=\frac1{2\pi \ii}\int_{C-\ii\infty}^{C+\ii\infty}
\Gamma\!\left(s-\tfrac12\right)\epsilon^{1/2-s}
E(\tau,s)\,ds,\qquad C>1.
\label{eq:mellin-EF}
\end{equation}
For $C>1$, absolute convergence and the decay of the gamma
function justify exchanging the sum and integral.

The sum $E(\tau,s)$ is an Eisenstein series. We start from reviewing the computation of its residue. It will be useful to consider the auxiliary sum over all nonzero integer pairs, including pairs with common divisors. Each such pair can be written as $(m,n)=k(r,d)$, where $k=\gcd(m,n)>0$ and $(r,d)=1$. Its contribution is
\[
\left(\frac{\tau_2}{|m\tau+n|^2}\right)^s
=k^{-2s}\left(\frac{\tau_2}{|r\tau+d|^2}\right)^s.
\]
Summing over all common divisors gives
$\sum_{k\ge1}k^{-2s}=\zeta(2s)$. This auxiliary sum also counts $(r,d)$ and $(-r,-d)$ separately, whereas $E(\tau,s)$ counts them once. Thus
\begin{equation}
\sum_{(m,n)\ne(0,0)}
\left(\frac{\tau_2}{|m\tau+n|^2}\right)^s
=2\zeta(2s)E(\tau,s).
\label{eq:gcd-full}
\end{equation}
Appendix~\ref{app:disk} shows that (\ref{eq:gcd-full}) has residue $\pi$, so we have
\begin{equation} \label{eq:E-res}
    \Res_{s=1}E(\tau,s) = \frac{\pi}{2\zeta(2)}=\frac3\pi.
\end{equation}
Moving the contour past $s=1$ therefore adds
\[
\Res_{s=1}\!\left[
\Gamma\!\left(s-\tfrac12\right)\epsilon^{1/2-s}E(\tau,s)
\right]
=\Gamma(\tfrac12)\epsilon^{-1/2}\frac3\pi
=\frac3{\sqrt{\pi\epsilon}}.
\]
This gives the leading divergence in $F(\tau,\epsilon)$: the $s=1$ residue fixes its coefficient, and the position of the pole fixes the power $\epsilon^{-1/2}$.

Multiplying by the factor $2/(\sqrt{\tau_2}|\eta(\tau)|^2)$
set aside above gives the divergence of the full bosonic sum:
\[
\frac2{\sqrt{\tau_2}\,|\eta(\tau)|^2}
\frac3{\sqrt{\pi\epsilon}}
=\frac6{\sqrt{\pi\epsilon}\,\sqrt{\tau_2}|\eta(\tau)|^2}.
\]
Choosing the cusp coefficient $-6/\sqrt{\pi\epsilon}$ cancels
this divergence, giving the subtraction in \eqref{eq:bosonic-cutoff}.

After subtracting the divergence that arises from the Eisenstein pole at $s=1$, the finite part of $F(\tau,\epsilon)$ is determined by the behavior of the integrand in Eq.~\eqref{eq:mellin-EF} at $s=\tfrac12$, for which the gamma factor has a pole with residue one and $\epsilon^{1/2-s}=1$. The finite part of $F(\tau,\epsilon)$ is then given by $E(\tau,\frac12)$, and since the full lattice sum in \eqref{eq:gcd-full} is regular at $s=\frac12$ after analytic continuation while $\zeta(2s)$ has a pole, we have $E(\tau,\frac12)=0$. This zero cancels the pole of the gamma function, such that the remaining contour integral vanishes as $\epsilon\to0$ \cite[Appendix B]{Stanford:2025llj}, so
\begin{equation}
    \lim_{\epsilon\to0} \left[F(\tau,\epsilon)-\frac{3}{\sqrt{\pi\epsilon}}\right]=E(\tau,\tfrac12)=0.
\end{equation}
Restoring the convergent terms then gives the same full bosonic partition function in both prescriptions. We next impose the spin restrictions.

\subsection{Spin restrictions and the cusp counterterm}
\label{sec:spin-counterterm}

The cusp divergence comes from the accumulation of solid-torus fillings, so restricting the allowed contractible cycles changes its coefficient. We can determine this change by organizing the primitive cycles into three winding-parity classes. As we show below, each class contributes equally, and each even spin structure allows precisely two of them.

A primitive pair $(r,d)$ cannot have
both entries even, so there are three possible winding parities:
\begin{equation} \label{eq:parity-class}
r\ \text{even},\ d\ \text{odd};\qquad
r\ \text{odd},\ d\ \text{even};\qquad
r\ \text{odd},\ d\ \text{odd}.
\end{equation}
Each of the three even spin sectors in \eqref{eq:sectors}
includes two of these classes. Without loss of generality, we choose one class from \eqref{eq:parity-class} and define
\begin{equation}
\begin{split}
E_{\mathrm{parity}}(\tau,s)
&=\sum_{\substack{(r,d)=1\\\text{chosen parities}}}
\tau_2^s|_\gamma,\\
F_{\mathrm{parity}}(\tau,\epsilon)
&=\sum_{\substack{(r,d)=1\\\text{chosen parities}}}
\left(\sqrt{\tau_2}\,e^{-\epsilon/\tau_2}\right)|_\gamma ,
\end{split}
\label{eq:parity-EF}
\end{equation}
where the subscript denotes one parity class, not a full spin sector. As before, $E_{\mathrm{parity}}(\tau,s)$ is the sum used in zeta-function regularization and $F_{\mathrm{parity}}(\tau,\epsilon)$ is the cutoff sum used in cusp regularization. Mellin inversion applies term by term, so
\begin{equation}
F_{\mathrm{parity}}(\tau,\epsilon)
=\frac{1}{2\pi\ii}
\int_{C-i\infty}^{C+i\infty}
\Gamma\!\left(s-\tfrac12\right)\epsilon^{1/2-s}
E_{\mathrm{parity}}(\tau,s)\,ds,\qquad C>1.
\label{eq:mellin-parity}
\end{equation}
The same Mellin formula applies to the restricted sum.
To find the leading divergence of $F_{\mathrm{parity}}(\tau,\epsilon)$, we
therefore need $\Res_{s=1}E_{\mathrm{parity}}(\tau,s)$.

We now compute the analogue of \eqref{eq:E-res} for a single parity class, for which we start instead with integer pairs $(m,n)$ with chosen parity. Fixing both parities doubles each lattice spacing, giving one quarter of the density and residue $\pi/4$ by the same counting argument as explained in Appendix~\ref{app:disk}. Now at least one entry is odd, so $k$ must be odd, and dividing by $k$ preserves both parities. The common-divisor factor becomes
$\sum_{k\ {\rm odd}}k^{-2s}=(1-2^{-2s})\zeta(2s)$:
\begin{equation}
\sum_{\substack{m,n\\\text{chosen parities}}}
\left(\frac{\tau_2}{|m\tau+n|^2}\right)^s
=2\zeta(2s)(1-2^{-2s})E_{\mathrm{parity}}(\tau,s).
\label{eq:gcd-parity}
\end{equation}
Dividing out these factors at $s=1$ gives
\begin{equation}
\Res_{s=1}E_{\mathrm{parity}}(\tau,s)
=\frac{\pi/4}{2\zeta(2)(1-2^{-2})}
=\frac1\pi.
\label{eq:parity-res}
\end{equation}
This holds for all three parity choices.

The Mellin formula \eqref{eq:mellin-parity} converts the
residue $1/\pi$ into a cutoff contribution from each parity class:
\begin{equation}
\Gamma\!\left(\tfrac12\right)\epsilon^{-1/2}\frac1\pi
=\frac1{\sqrt{\pi\epsilon}}.
\label{eq:parity-cutoff}
\end{equation}
Scalar corrections are absolutely summable, and the
phase-dependent terms have no $s=1$ pole
(see Appendix~\ref{app:cusp-pole}). The leading divergence of
the spin-restricted sum is therefore
\begin{equation}
\underbrace{2}_{\text{scalar vacuum terms}}\;\times\;
\underbrace{2}_{\text{allowed parity classes}}\;\times\;
\underbrace{\frac1{\sqrt{\pi\epsilon}}}_{\text{one class}}
=\frac4{\sqrt{\pi\epsilon}}.
\label{eq:four}
\end{equation}
The cusp contribution must cancel this
divergence, so
\begin{equation}
Z_{\rm cusp}^{\text{even spin}}(\tau)
=-\frac4{\sqrt{\pi\epsilon}\,\sqrt{\tau_2}\,|\eta(\tau)|^2},
\qquad \text{for }R^-,NS^+,NS^-.
\label{eq:counterterm}
\end{equation}
Adding the cusp contribution gives the cutoff expression
\begin{equation}
Z_{\s}(\tau) = \frac1{\sqrt{\tau_2}\,|\eta(\tau)|^2}
\left[
-\frac4{\sqrt{\pi\epsilon}}
+\sum_{\substack{(r,d)=1\\\text{spin-compatible}}}
\left(\sqrt{\tau_2}\,
|q|^{-(c-1)/12}|1-q|^2e^{-\epsilon/\tau_2}\right)\bigg|_\gamma
\right].
\label{eq:spin-cutoff}
\end{equation}
This is the subtraction in \eqref{eq:main-result}, with a
common coefficient for the three full spin-sector traces.

We have determined the cusp subtraction. We now ask whether the result left after this subtraction agrees with zeta-function regularization. For the parity sums, we show in Appendix~\ref{app:reduction} that
\begin{equation}
F(\tau,\epsilon)=\frac3{\sqrt{\pi\epsilon}}+o(1),
\qquad
F_{\mathrm{parity}}(\tau,\epsilon)
=\frac1{\sqrt{\pi\epsilon}}+o(1),
\label{eq:F-asymp}
\end{equation}
and that $E_{\mathrm{parity}}(\tau,\tfrac12)=0$, so
zeta-function and cusp regularization agree for each spin sum:
\begin{equation}
\lim_{\epsilon\to0}
\left[F_{\mathrm{parity}}(\tau,\epsilon)
-\frac1{\sqrt{\pi\epsilon}}\right]
=E_{\mathrm{parity}}(\tau,\tfrac12)=0.
\label{eq:basic-agreement}
\end{equation}
This finite agreement uses the full sum \eqref{eq:parity-EF} not just the $s=1$ residue. 

Appendix~\ref{app:cusp-pole} derives the same pole and counterterm from the energy density; we should note that this is a different problem than the finite negative threshold weights which we review in Appendix~\ref{app:threshold}.

\section{Unequal left- and right-moving central charges}
\label{sec:unequal-central-charges}

So far we have assumed $c_L=c_R$. When the left- and right-moving central charges differ, the bulk gravitational Chern-Simons term induces a gravitational anomaly in the boundary theory \cite{Kraus:2005zm}. In the black hole partition function, this anomaly appears as relative phases between different fillings \cite{Boruch:2026hbr}. We ask how these phases change the leading divergence and the required cusp subtraction. Section~\ref{subsec:bos} treats bosonic gravity. Section~\ref{subsec:spin-anomaly} determines the spin-dependent anomaly phase and applies it to pure gravity with spin structures and to $\mathcal N=(1,1)$ supergravity.

\subsection{Bosonic gravity} \label{subsec:bos}
Let $\Delta c=c_L-c_R$. In bosonic gravity, the partition function is given by \cite{Boruch:2026hbr},
\begin{equation} \label{eq:Z-bos-anomaly}
    Z_{\rm bos}(\tau) = \sum_{\gamma \in \mathbb{Z} \backslash PSL(2,\Z)} e^{\frac{\pi\ii\Delta c}{12} \Phi(\gamma)}\, \chi^L_{\text{vac}}(\gamma \tau)\bar{\chi}^R_{\text{vac}}(\gamma \bar{\tau})\,,
\end{equation}
where $\Phi(\gamma)$ is the Rademacher function.\footnote{The Rademacher function is defined by $\Phi(\gamma) = \frac{a+d}{r} - 12 \, \text{sgn}(r)\, {\sf s}(d,|r|)$ for $r\neq 0$ or $b/d$ for $r=0$, where ${\sf s}(d,r)=\frac{1}{4r}\sum_{k=1}^{r-1} \cot \frac{\pi k}{r} \cot \frac{\pi d k}{r}$ for $r>0$ is the Dedekind sum. It satisfies the identity $\Phi(\gamma_1 \gamma_2) = \Phi(\gamma_1) + \Phi(\gamma_2) - 3 \, \text{sgn}(r_1 r_2 r_{12})$ where $r_1,r_2,r_{12}$ are the bottom left entries of $\gamma_1,\gamma_2,\gamma_1\gamma_2$, respectively (see e.g. \cite{kirby1994, jeffrey2010}).} The role of $\Phi(\gamma)$ is to compensate the change of the vacuum amplitude under a change of representative $\gamma\mapsto T\gamma$ and in the bulk theory it arises from coupling pure 3D gravity to a gravitational Chern-Simons term \cite{Boruch:2026hbr}.
The allowed central-charge differences satisfy \cite{Kapustin:2014dxa}
\begin{equation}
    \Delta c\in8\Z\,.
\end{equation}
The vacuum characters with left- and right-moving central charges $c_L$ and $c_R$ are given by
\begin{equation} \label{eq:characters}
    \chi_{\text{vac}}^L(\tau) = \frac{(1-q)q^{-(c_L-1)/24}}{\eta(\tau)}\,,\qquad \chi_{\text{vac}}^R(\tau) = \frac{(1-q)q^{-(c_R-1)/24}}{\eta(\tau)}\,.
\end{equation}

Using the vertical bar notation of \eqref{eq:slash}, write
\begin{equation}
    \gamma\tau = \tau_1|_\gamma+\ii\tau_2|_\gamma\,,\qquad \tau_2|_\gamma = \frac{\tau_2}{|r\tau+d|^2}\,,\qquad \alpha = \frac{c_L+c_R-2}{24}\,.
\end{equation}
The summand in~\eqref{eq:Z-bos-anomaly} is then
\begin{equation} \label{eq:anomaly-filling}
    \mathcal A_\gamma(\tau) = \frac{e^{\frac{\pi\ii\Delta c}{12}\Phi(\gamma)}}{\sqrt{\tau_2}|\eta(\tau)|^2}
    \left(\sqrt{\tau_2}\,e^{2\pi\alpha\tau_2}e^{-2\pi\ii\Delta c\tau_1/24}(1-q)(1-\bar q)\right)|_\gamma\,.
\end{equation}
As in Section~\ref{sec:bosonic-regularization}, the vertical bar evaluates every $\tau$-dependent factor inside the parentheses at $\gamma\tau$, including $q$ and $\bar q$.
We regulate the black hole sum \eqref{eq:Z-bos-anomaly} in two ways: by introducing the parameter $s$ for zeta-function regularization, or by imposing the cutoff $\epsilon$ for cusp regularization:
\begin{equation} \label{eq:anomaly-regulators}
\begin{aligned}
    Z_{\rm bos}^{\mathrm{zeta}}(\tau;s)&=\sum_\gamma (\tau_2|_\gamma)^{s-1/2}\mathcal A_\gamma(\tau)\,,\\
    Z_{\rm bos}^{\mathrm{cutoff}}(\tau;\epsilon)&=\sum_\gamma e^{-\epsilon/(\tau_2|_\gamma)}\mathcal A_\gamma(\tau)\,.
\end{aligned}
\end{equation}
The first sum converges absolutely for $\Re s>1$ and is analytically continued to $s=\tfrac12$; the second converges for $\epsilon>0$ and does not yet include the cusp contribution. Just as in~\eqref{eq:mellin-EF}, their exact relation is
\begin{equation} \label{eq:anomaly-mellin}
    Z_{\rm bos}^{\mathrm{cutoff}}(\tau;\epsilon)=\frac1{2\pi\ii}\int_{C-\ii\infty}^{C+\ii\infty} \Gamma\!\left(s-\tfrac12\right)\epsilon^{1/2-s}Z_{\rm bos}^{\mathrm{zeta}}(\tau;s)\,ds\,,\qquad C>1\,.
\end{equation}
The $s=1$ Eisenstein pole gives the leading divergence of the cutoff sum as $\epsilon\to0$:
\begin{equation} \label{eq:cL-cR-div}
    Z_{\rm bos}^{\mathrm{cutoff}}(\tau;\epsilon)
    =\frac{3\left(2\delta_{\Delta c,0}-\delta_{\Delta c,24}-\delta_{\Delta c,-24}\right)}{\sqrt{\pi\epsilon\tau_2}\,|\eta(\tau)|^2}
    +o(\epsilon^{-1/2})\,.
\end{equation}
Here $o(\epsilon^{-1/2})$ denotes terms smaller than $\epsilon^{-1/2}$, not necessarily finite terms. The cusp contribution cancels the displayed term, which is nonzero only for $\Delta c=0,\pm24$.

An explicit derivation of~\eqref{eq:cL-cR-div} is given in Appendix~\ref{app:anomalous-pole}, but its physical origin can be understood as follows. Just as in Section~\ref{sec:mellin-divergence}, only the leading behavior at $\tau_2|_\gamma\to0$ can contribute to this pole. In this limit,
\begin{equation} \label{eq:anomaly-cusp-expansion}
    \mathcal A_\gamma(\tau) = \frac{e^{\frac{\pi\ii\Delta c}{12} \Phi(\gamma)}}{\sqrt{\tau_2}|\eta(\tau)|^2}
    \left(\sqrt{\tau_2}\left[2e^{-2\pi\ii \Delta c\tau_1/24}-e^{2\pi\ii(1-\Delta c/24)\tau_1}
    -e^{2\pi\ii(-1-\Delta c/24)\tau_1}+O(\tau_2)\right]\right)|_\gamma\,,
\end{equation}
where the $O(\tau_2)|_\gamma$ terms are summable at $s=1$, so they cannot affect the leading pole. The four vacuum terms are the same as those listed in Appendix~\ref{app:energy}, Eq.~\eqref{eq:four-seeds}. Only terms with no remaining angular dependence contribute to the $s=1$ pole, as shown in Appendix~\ref{app:anomalous-pole}. In \eqref{eq:anomaly-cusp-expansion}, this occurs at $\Delta c=0,24,-24$, respectively, giving the following table:
\begin{equation}
\begin{array}{c|c|c}
\Delta c & \text{vacuum terms with no remaining angular phase} & \text{total coefficient}\\ \hline
0 & 1+q\bar q & 2\\
24 & -q & -1\\
-24 & -\bar q & -1
\end{array}
\label{eq:anomaly-vacuum-terms}
\end{equation}
where for example, at $\Delta c=24$ the factor $e^{-2\pi\ii\tau_1}$ cancels the angular phase of $-q$, leaving $(-e^{-2\pi\tau_2})|_\gamma\to-1$. The table therefore gives
\begin{equation} \label{eq:cL-cR-deltas}
   2\delta_{\Delta c,0}-\delta_{\Delta c,24}-\delta_{\Delta c,-24}\,.
\end{equation}
Each nonoscillating term multiplies the same basic Eisenstein series, with residue $3/\pi$ at $s=1$, which becomes $3/\sqrt{\pi\epsilon}$ in the smoothly cut-off sum after applying the Mellin transform~\eqref{eq:mellin-EF}. By restoring the Eisenstein residue and the common factor $\frac{1}{\sqrt{\tau_2}|\eta(\tau)|^2}$ in~\eqref{eq:cL-cR-deltas}, we recover Eq.~\eqref{eq:cL-cR-div}.
The required cusp contribution has the same dependence on $\tau$, with the opposite coefficient, and cancels this leading divergence.

\subsection{Gravity with spin structures} \label{subsec:spin-anomaly}
With a spin structure, the gravitational anomaly can take half-integer values \cite{Kapustin:2014dxa},
\begin{equation}
    \Delta c=c_L-c_R\in\tfrac12\Z\,.
\end{equation}
For a fixed even spin structure $\s$, the sum over compatible solid-torus fillings takes the form \cite{Boruch:2026hbr}
\begin{equation} \label{eq:Z-spin-anom}
    Z_{\s}(\tau) = \sum_{\gamma \in \mathbb{Z} \backslash \Gamma_{\s}} e^{\frac{\pi\ii\Delta c}{12} \Phi_{\s}(\gamma)}\, \chi^L_{\s}(\gamma \tau)\bar{\chi}^R_{\s}(\gamma \bar{\tau})\,,
\end{equation}
where $\Gamma_{\s}$ is given by~\eqref{eq:spin-Gamma}, and the seed functions $\chi_{\s}(\tau)$ depend on the theory and the spin structure. We now determine $\Phi_{\s}$ in terms of the ordinary Rademacher function $\Phi$. We first compute the modular phase of one chiral Majorana fermion, then obtain the compensating filling phase for general $\Delta c$. A bulk derivation is provided in Appendix~\ref{app:spin-anomaly}.

The pure gravitational anomaly depends on $\Delta c$, not on the
detailed field content. A left-moving Majorana fermion, with
$\Delta c=\tfrac12$, therefore provides a simple reference.
We use only its anomaly phase; we do not add fermions to the bulk
theory or replace the vacuum characters $\chi_{\s}$.
Its partition functions for the three even spin structures are
\begin{equation} \label{eq:phase-f}
    f_{NS^+}(\tau) = \frac{\eta(\tau/2)}{\eta(\tau)}\,,\qquad f_{NS^-}(\tau) = e^{-\ii\pi/24} \frac{\eta((\tau+1)/2)}{\eta(\tau)}\,,\qquad f_{R^-}(\tau) = \sqrt{2}\frac{\eta(2\tau)}{\eta(\tau)}\,.
\end{equation}
Each is a ratio of eta functions. The numerator and denominator eta functions transform with different modular matrices, so their ratio acquires a phase determined by the difference of the corresponding Rademacher functions.

Let $\gamma=\left(\begin{smallmatrix}a&b\\r&d\end{smallmatrix}\right)
\in\Gamma_{\s}$ preserve the chosen spin structure.
The denominator has argument $\gamma\tau$. To transform the
numerator, define $\gamma_{\s}$ by its action on the corresponding
argument:
\begin{equation}
    \frac{\gamma\tau}{2}
    =\gamma_{NS^+}\!\left(\frac\tau2\right),\qquad
    \frac{\gamma\tau+1}{2}
    =\gamma_{NS^-}\!\left(\frac{\tau+1}{2}\right),\qquad
    2\gamma\tau=\gamma_{R^-}(2\tau).
\label{eq:spin-eta-arguments}
\end{equation}
These identities give
\begin{equation}
    \gamma_{NS^+}= \begin{pmatrix} a&b/2\\ 2r&d \end{pmatrix}\,, \qquad \gamma_{NS^-}= \begin{pmatrix} a+r&(b+d-a-r)/2\\ 2r&d-r \end{pmatrix}\,, \qquad \gamma_{R^-}= \begin{pmatrix} a&2b\\ r/2&d \end{pmatrix}
\label{eq:spin-eta-matrices}
\end{equation}
in the respective sectors. The spin restriction ensures that each
matrix has integer entries and belongs to $SL(2,\Z)$.

For example, in $NS^+$ the numerator transforms with
$\gamma_{NS^+}$ and the denominator with $\gamma$.
Their modular-weight factors are identical, since
$2r(\tau/2)+d=r\tau+d$, and cancel in the ratio. Hence
\begin{equation}
    \frac{\eta((\gamma\tau)/2)}{\eta(\gamma\tau)}
    =e^{\frac{\ii\pi}{12}[\Phi(\gamma_{NS^+})-\Phi(\gamma)]}
     \frac{\eta(\tau/2)}{\eta(\tau)}.
\label{eq:NS-eta-phase}
\end{equation}
The same cancellation occurs in the other two sectors, giving
\begin{equation}
    f_{\s}(\gamma\tau) = e^{\frac{\ii\pi}{12} [\Phi(\gamma_{\s})-\Phi(\gamma)]} f_{\s}(\tau)\,.
\label{eq:Majorana-modular-phase}
\end{equation}

Since one left-moving Majorana contributes $\Delta c=\tfrac12$,
the multiplier in \eqref{eq:Majorana-modular-phase} is raised to
$2\Delta c$.
The compensating filling phase is its inverse,
$[f_{\s}(\gamma\tau)/f_{\s}(\tau)]^{-2\Delta c}
=\exp\{\pi\ii\Delta c[\Phi(\gamma)-\Phi(\gamma_{\s})]/6\}$.
Comparing with the phase in \eqref{eq:Z-spin-anom} gives
\begin{equation} \label{eq:spin-anomaly-phase}
    \Phi_{\s}(\gamma) = 2\left[\Phi(\gamma)-\Phi(\gamma_{\s})\right]\,.
\end{equation}
Since the Rademacher function is integer-valued, $\Phi_{\mathfrak{s}}(\gamma)\in 2\mathbb Z$, so the anomalous phase can become trivial already for $\Delta c\in12\mathbb Z$, rather than requiring $\Delta c\in24\mathbb Z$. This allows a leading cusp divergence at $\Delta c=\pm12$ when the vacuum seed contains half-integer angular harmonics, as we will see happens for $\mathcal N=(1,1)$ supergravity.

\subsubsection*{Pure gravity with spin structures}
We first consider pure gravity with a specified spin structure but no gravitino. The local vacuum determinant is unchanged, while the spin structure restricts the allowed fillings and determines the anomaly phase. The partition function is given by~\eqref{eq:Z-spin-anom}, with $\chi_{NS^\pm}(\tau)=\chi_{\rm vac}(\tau)$ and $\chi_{R^-}(\tau)=\chi_{\rm vac}(S\tau)$.

The reduced vacuum seed has the same angular terms as \eqref{eq:anomaly-cusp-expansion}. Including the spin-anomaly phase, the calculation in Appendix~\ref{app:anomalous-pole} gives the same selection $\Delta c=0,\pm24$, with residue $2/\pi$ for each contributing term. Thus for each of the three even spin structures,\footnote{The $R^-$ result follows from the $NS^+$ result by exchanging the time and space cycles, $\tau\mapsto-1/\tau$. The cusp factor $1/(\sqrt{\tau_2}|\eta(\tau)|^2)$ is unchanged under this transformation, so the leading divergence has the same coefficient.}
\begin{equation} \label{eq:cL-cR-div-ferm}
    Z_{\s}^{\mathrm{cutoff}}(\tau;\epsilon)
    =\frac{2\left(2\delta_{\Delta c,0}-\delta_{\Delta c,24}-\delta_{\Delta c,-24}\right)}{\sqrt{\pi\epsilon\tau_2}\,|\eta(\tau)|^2}
    +o(\epsilon^{-1/2})\,,
\end{equation}
where $Z_{\s}^{\mathrm{cutoff}}$ denotes the spin-restricted sum with the same exponential cutoff as in \eqref{eq:anomaly-regulators}. Its leading term reproduces \eqref{eq:four} for $c_L=c_R$. Hence just as in the bosonic case, the leading cusp divergence only has nonvanishing contributions from $c_L-c_R=0,\pm 24$.

\subsubsection*{$\N=(1,1)$ supergravity}
\label{subsec:1,1}
We next include one left- and one right-moving gravitino and denote the partition functions by $Z_{\s}^{(1,1)}$ to distinguish them from those of pure gravity. The main change is that the $\N=1$ vacuum has a null state already at level $1/2$, generated by the supercurrent $G_{-1/2}$, with $G_{-1/2}^2=L_{-1}$. After the universal NS descendants are extracted, the remaining vacuum quotient is therefore $1\mp q^{1/2}$, rather than the Virasoro factor $1-q$.

We begin with the $NS^+$ sector, whose spin structure is preserved by $\Gamma^0(2)$. The sum $Z_{NS^+}^{(1,1)}$ has the form~\eqref{eq:Z-spin-anom}, with vacuum characters
\begin{equation}
    \chi^L_{NS^+}(\tau) = q^{-(c_L-\frac32)/24} \frac{\eta(\tau/2)}{\eta(\tau)^2} \left(1+q^{1/2}\right)\,,\qquad \chi^R_{NS^+}(\tau) = q^{-(c_R-\frac32)/24} \frac{\eta(\tau/2)}{\eta(\tau)^2} \left(1+q^{1/2}\right)\,.
\end{equation}
The $NS^-$ and $R^-$ partition functions are given by
\begin{equation} \label{eq:N=1relations}
    Z_{NS^-}^{(1,1)}(\tau) = e^{\pi\ii\Delta c/12}Z_{NS^+}^{(1,1)}(\tau+1)\,,\qquad Z_{R^-}^{(1,1)}(\tau) = Z_{NS^+}^{(1,1)}(-1/\tau)\,.
\end{equation}

It is enough to determine the $s=1$ Eisenstein pole in the $NS^+$ sector. The supersymmetric vacuum-energy shifts are identical on the two sides and therefore cancel from the angular momentum. In the $\tau_2|_\gamma\to0$ limit, the summand for the $NS^+$ partition function is given by
\begin{equation}
    \mathcal A_\gamma^{NS^+}(\tau) = \frac{|\vartheta_{4}(\tau)|\,e^{\frac{\pi\ii\Delta c}{12} \Phi_{NS^+}(\gamma)}}{\sqrt{\tau_2}|\eta(\tau)|^3} \left(\sqrt{\tau_2}\left[2e^{-2\pi\ii\Delta c\tau_1/24}+e^{2\pi\ii(\frac12-\frac{\Delta c}{24})\tau_1} + e^{2\pi\ii(-\frac12-\frac{\Delta c}{24})\tau_1}+O(\tau_2)\right]\right)|_\gamma\,.
\end{equation}
As in the nonsupersymmetric cases, only terms with no remaining angular dependence contribute to the $s=1$ pole. Here this occurs at $\Delta c=0,\pm12$. The vacuum-term coefficients, again before including the residue and determinant, are
\begin{equation}
\begin{array}{c|c|c|c}
    \Delta c & \text{contributing terms in }NS^+ & NS^+\text{ coefficient} & NS^-\text{ coefficient}\\ \hline
    0 & 1+(q\bar q)^{1/2} & 2 & 2\\
    12 & q^{1/2} & 1 & -1\\
    -12 & \bar q^{1/2} & 1 & -1
\end{array}
\label{eq:N11-vacuum-terms}
\end{equation}
where the modular relation~\eqref{eq:N=1relations} supplies a minus sign at $\Delta c=\pm 12$, and none at $\Delta c=0$ for the $NS^-$ coefficients. The $R^-$ result follows from an $S$-transform of the full $NS^+$ residue, including its cusp determinant; in particular it uses the same selection rule and coefficients as $NS^+$.

The $\tau$-dependent factor in the $NS^+$ cusp amplitude was computed in \cite{Stanford:2025llj} and is given by $|\eta(\tau/2)|^2/(\sqrt{\tau_2}|\eta(\tau)|^4)$. The corresponding factors in the other two spin structures follow by $T$- and $S$-transforms, using
\begin{equation}
    \eta(\tau+1) = e^{\ii\pi/12}\eta(\tau)\,,\qquad \eta(-1/\tau) = \sqrt{-\ii\tau}\,\eta(\tau)\,,\qquad \mathrm{Im}(-1/\tau)=\tau_2/|\tau|^2\,.
\end{equation}
Finally, the regulated cusp one-loop determinant was found in \cite{Stanford:2025llj} to have a leading logarithmic divergence $\frac12\log(1/\epsilon)$. Combining the spin-dependent cusp factors with the Eisenstein residue $2/\pi$ and the anomaly-dependent coefficients of~\eqref{eq:N11-vacuum-terms} gives
\begin{equation} \label{eq:cL-cR-div-N11}
\begin{aligned}
    Z_{\s}^{(1,1),\,\mathrm{cutoff}}(\tau;\epsilon) &= \frac{2}{\sqrt{\pi\epsilon\tau_2}\,|\eta(\tau)|^4}\\
    &\quad\times\begin{cases}
        \left(2\delta_{\Delta c,0}-\delta_{\Delta c,12}-\delta_{\Delta c,-12}\right) ~\left|\eta\left(\frac{\tau+1}{2}\right)\right|^2\quad & \s=NS- \\
        \left(2\delta_{\Delta c,0}+\delta_{\Delta c,12}+\delta_{\Delta c,-12}\right) ~\left|\eta\left(\frac{\tau}{2}\right)\right|^2 & \s=NS+\\
        \left(2\delta_{\Delta c,0}+\delta_{\Delta c,12}+\delta_{\Delta c,-12}\right) ~2|\eta(2\tau)|^2 & \s=R-
    \end{cases}\\
    &\quad+o(\epsilon^{-1/2})\,.
\end{aligned}
\end{equation}
The displayed leading term is nonzero only for $c_L-c_R=0,\pm12$.

The three results~\eqref{eq:cL-cR-div},~\eqref{eq:cL-cR-div-ferm}, and~\eqref{eq:cL-cR-div-N11} have a similar structure: the Eisenstein residue fixes the universal size of the accumulation divergence, while the reduced vacuum seed determines which anomalous angular-momentum shift can reach the nonoscillating channel. The bosonic and pure fermionic vacuum seeds contain $J=0,\pm1$, giving $\Delta c=0,\pm24$, whereas the $\mathcal N=(1,1)$ seed contains $J=0,\pm\tfrac12$, giving $\Delta c=0,\pm12$. The relative signs between spin structures are then fixed by modular covariance.

\section{Discussion}

\paragraph{Relationship to the Eisensteins in BTT.}
The restriction to spin structure compatible fillings also has a simple interpretation in terms of the Eisenstein series studied in \cite{Boruch:2026hbr}.\footnote{See also \cite{Benjamin:2020zbs, Benjamin:2021ygh, DiUbaldo:2023qli, Boruch:2025ilr} for related applications of modular and spectral methods to two-dimensional CFTs and AdS$_3$ gravity.} For each even spin structure, the corresponding modular group has a fundamental domain with two inequivalent cusps, of widths two and one, referred to as NS and R cusps, respectively, in \cite{Witten:2007kt}. The width-two cusp corresponds to cycles with antiperiodic fermion boundary conditions, which can contract smoothly in the bulk. The sum over these fillings is precisely the Eisenstein series associated with this cusp, up to a normalization factor:
\begin{equation}
\begin{split}
    E_{NS^+}(\tau,s) &= 2^s E_{\infty}^{\Gamma^0(2)}(\tau,s)\,,\quad\\ E_{NS^-}(\tau,s) &= 2^s E_{\infty}^{\Gamma_\theta}(\tau,s)\,,\quad\\ E_{R^-}(\tau,s) &= 2^s E_{0}^{\Gamma_0(2)}(\tau,s)\,,
\end{split}
\end{equation}
where the left-hand side are the Eisenstein series considered in Eq.~\eqref{eq:parity-EF} while the right-hand side are the Eisenstein series associated to cusps of the fundamental domain in BTT. The factor $2^s$ accounts for the cusp width, and since the fundamental domain is three times larger in the presence of spin structure, this leads to an $s=1$ residue of $2/3$ times the bosonic one, i.e. given by $2/\pi$ rather than $3/\pi$. In particular, the width-two cusp accounts for two thirds of the primitive filling cycles, which is why the coefficient of the divergence in the modular sum is the same in all three even spin sectors.\footnote{Similar geometric arguments apply to the case of $\mathcal N=(1,1)$ supergravity, but there the gravitino determinant introduces nontrivial dependence on the spin structure.} The Eisenstein series associated with the other, width-one cusp also enters the spectral decomposition in \cite{Boruch:2026hbr}, but does not appear in our sum over smooth spin fillings.

\paragraph{Cases where no cusp contribution is needed.}
Our analysis also shows that a cusp geometry can exist even when no counterterm is needed. When the gravitational anomaly is nonzero, the relative phases between smooth fillings can remove the leading divergence without introducing an additional geometric contribution. In pure gravity, with or without a spin structure, this occurs for $\Delta c\neq 0,\pm24$, while in $\mathcal N=(1,1)$ supergravity this occurs for $\Delta c\neq 0,\pm12$. In these cases, no vacuum-character term has the angular dependence needed to cancel the anomalous shift, and the $s=1$ Eisenstein residue vanishes.

For $\Delta c\in24\Z$ in bosonic gravity, or $\Delta c\in12\Z$ in the presence of a spin structure, the anomaly phase is trivial and the cusp geometry has the same one-loop determinant with the same $\tau$ and $\epsilon$ dependence as when $\Delta c=0$. When the $s=1$ Eisenstein residue vanishes, the cusp fugacity is then left unfixed, and a natural choice might be to set it to zero. Similarly, the purely periodic $R^+$ spin structure admits no smooth spin fillings and hence no accumulation divergence, yet admits a cusp geometry with an unfixed fugacity.

\paragraph{Comments on periodic boundary conditions.}
The fully periodic spin structure is particularly interesting because the limiting cusp geometry can carry a spin structure that does not admit any compact smooth filling. Indeed, the $R^+$ boundary torus represents a nontrivial element of the spin bordism group, so it cannot bound a smooth spin three-manifold (see e.g. discussions in \cite{Maloney:2007ud, Boruch:2026hbr}). However, it can extend over the noncompact cusp geometry, in which no boundary cycle contracts. Without supersymmetry, the local cusp one-loop determinant has the same $\tau$ and $\epsilon$ dependence for any choice of spin structure, and in particular it still diverges for the $R^+$ spin structure, despite there being no accumulation divergence to cancel. Requiring a finite partition function then imposes that its fugacity be set to zero.

In contrast, in $\mathcal N=(1,1)$ supergravity, Stanford and Yan \cite{Stanford:2025llj} found that the graviton and gravitino one-loop determinants cancel for the fully periodic spin structure, giving a finite, $\tau$-independent cusp contribution. Assigning this cusp a nonzero fugacity can then produce a nonvanishing Ramond index, despite there being no smooth fillings of the torus. In the presence of a gravitational anomaly, the local determinant cancellation is expected to persist, but the cusp contribution must also satisfy the anomalous modular transformation laws. In particular, a nonzero constant Ramond index is compatible with these transformations only when $\Delta c\in12\mathbb Z$. In these cases, the cusp fugacity, and hence the Ramond index, are not fixed by the cancellation of the accumulation divergence.

\section*{Acknowledgments}
We want to thank Jan Boruch, Douglas Stanford, and Joaquin Turiaci for valuable discussions and suggestions. CY was supported in part by the Simons Investigator in Physics Award MP-SIP-0001737 and U.S. Department of Energy grant DE-SC0007870 and DE-SC0021013.  This work was performed in part at the Aspen Center for Physics, which is supported by National Science Foundation grant PHY-2210452. We also acknowledge assistance from OpenAI GPT-6 in manuscript revisions and mathematical derivation.

\appendix

\section{Details of the parity-restricted cusp sum}
\label{app:parity}

This appendix supplies the detailed proofs used in Section \ref{sec:tau}. We first derive the residues that determine the cusp subtraction in \ref{app:disk}. We then compute the finite part of $F_{\mathrm{parity}}(\tau,\epsilon)$ after subtracting its divergence and relate it to the zeta-regularized $E_{\mathrm{parity}}(\tau,s)$ in \ref{app:reduction}.

\subsection{The residue from counting lattice points in a disk}
\label{app:disk}
Here we justify the residues used in \eqref{eq:gcd-full} and \eqref{eq:gcd-parity} by counting integer lattice points. We first consider the sum over all nonzero integer pairs, whose asymptotic density is easy to determine, and then relate this sum to the Eisenstein by factoring out common divisors. Without loss of generality, fix any one of the three choices 
\begin{equation}\label{eq:app-choices}
    r\ \text{even},\ d\ \text{odd};\qquad
    r\ \text{odd},\ d\ \text{even};\qquad
    r\ \text{odd},\ d\ \text{odd}.
\end{equation}
The subscript ``parity'' denotes this fixed choice throughout the proof. Compare its auxiliary lattice sum with the full auxiliary sum:
\begin{equation} \label{eq:Zlattice}
\begin{split}
\mathcal Z_{\rm full}(\tau,s)
&=\sum_{(m,n)\ne(0,0)}
\left(\frac{\tau_2}{|m\tau+n|^2}\right)^s,\\
\mathcal Z_{\mathrm{parity}}(\tau,s)
&=\sum_{\substack{m,n\in\Z\\\text{chosen parities}}}
\left(\frac{\tau_2}{|m\tau+n|^2}\right)^s.
\end{split}
\end{equation}
Unlike the sums over gravitational fillings, these auxiliary sums include integer pairs with nontrivial common divisors and separately count $(m,n)$ and $(-m,-n)$.

Define the counting functions at the point where they are
needed:
\begin{equation} \label{eq:counts}
\begin{split}
N(R)&=\#\left\{(m,n)\in\Z^2\setminus\{(0,0)\}:
\frac{|m\tau+n|^2}{\tau_2}\le R^2\right\},\\
N_{\mathrm{parity}}(R)&=
\#\left\{(m,n)\in\Z^2:
\begin{gathered}
\text{chosen parities},\\
|m\tau+n|^2/\tau_2\le R^2
\end{gathered}\right\}.
\end{split}
\end{equation}
Change variables as follows:
\begin{equation}
x=\frac{m\tau_1+n}{\sqrt{\tau_2}},\quad
y=m\sqrt{\tau_2},\qquad
m=\frac{y}{\sqrt{\tau_2}},\quad
n=\sqrt{\tau_2}x-\frac{\tau_1y}{\sqrt{\tau_2}},
\label{eq:xy}
\end{equation}
which preserves area (the absolute Jacobian is one) and maps the ellipse in $(m,n)$ to a disk in $(x,y)$ of radius $R$. A fundamental cell of the transformed full lattice has area one, so the number of lattice points is approximately $\pi R^2$. Fixing a parity class doubles the lattice spacing in both integer directions, so its cells have area four. Only cells intersecting the boundary of the disk cause an error in the counting, and since the cells have bounded diameter at fixed $\tau$, such cells lie in an annulus of fixed width and area $O_\tau(R+1)$. Therefore
\begin{equation} \label{eq:count-error}
\begin{split}
N(R)&=\pi R^2+O_\tau(R+1),\\
N_{\mathrm{parity}}(R)
&=\frac\pi4R^2+O_\tau(R+1),
\end{split}
\end{equation}
where we will see that the coefficients of the $R^2$ terms determine the residue exactly while the subleading terms cannot affect the pole at $s=1$.

The corresponding continuous integral is particularly simple,
\begin{align}
\int_{\{|m\tau+n|^2/\tau_2\ge1\}}
\left(\frac{\tau_2}{|m\tau+n|^2}\right)^s\,dm\,dn
&=\int_{\{x^2+y^2\ge1\}}(x^2+y^2)^{-s}\,dx\,dy
\nonumber\\
&=2\pi\int_1^\infty R^{1-2s}\,dR
=\frac{\pi}{s-1}
\label{eq:disk-integral}
\end{align}
where $m$ and $n$ are real integration variables.
The residue of \eqref{eq:disk-integral} is the disk-area coefficient $\pi$.
We now explain why the discrete sum has exactly that residue.

For a lattice point at radius
$R_0=|m\tau+n|/\sqrt{\tau_2}$, write its contribution as
\[
R_0^{-2s}=2s\int_{R_0}^\infty R^{-2s-1}\,dR.
\]
When we sum these integrals, a point contributes at radius
$R$ precisely when $R_0\le R$, so the number of contributions
is $N(R)$. Separating the finitely many points with $R_0\le1$,
whose contributions are entire in $s$, gives for $\Re s>1$
\begin{align}
\mathcal Z_{\rm full}(\tau,s)
&=\text{entire terms}
+2s\int_1^\infty N(R)R^{-2s-1}\,dR\nonumber\\
&=\frac{\pi s}{s-1}
+\text{a function holomorphic near }s=1.
\label{eq:stieltjes}
\end{align}
The boundary term at $R=1$ is included in the entire terms.
Indeed, inserting the error in \eqref{eq:count-error} leaves
an integral bounded on compact subsets of $\Re s>\tfrac12$.
Replacing $N$ by $N_{\mathrm{parity}}$ gives
$\pi s/[4(s-1)]$ with the same holomorphic-error argument.
Thus
\begin{equation} \label{eq:lattice-res}
\begin{split}
\Res_{s=1}\mathcal Z_{\rm full}(\tau,s)&=\pi,\\
\Res_{s=1}\mathcal Z_{\mathrm{parity}}(\tau,s)
&=\frac\pi4.
\end{split}
\end{equation}
This residue is $\pi/4$ for each of the three choices \eqref{eq:app-choices}.

\subsection{An exact reduction to Stanford-Yan's length sum}
\label{app:reduction}
Here we justify \eqref{eq:basic-agreement}, proving agreement of the finite parts of the parity sums by reducing them to Stanford-Yan's unrestricted $E(\tau,s)$ and $F(\tau,\epsilon)$, whose cutoff and zeta-function regularizations are already known to agree.

We first write $E_{\mathrm{parity}}(\tau,s)$ in terms of the unrestricted $E(\tau,s)$ at different moduli. For example, consider the class with $r$ even and $d$ odd. Restricting the first integer to be even is equivalent to replacing $\tau$ by $2\tau$, while subtracting the pairs with both entries even removes the unwanted even-even class:
$$\mathcal Z_{r\,\mathrm{even},\,d\,\mathrm{odd}}(\tau,s) = 2^{-s}\mathcal Z_{\mathrm{full}}(2\tau,s) - 2^{-2s}\mathcal Z_{\mathrm{full}}(\tau,s).$$
Then by applying similar arguments to the other two classes and rearranging the sums in \eqref{eq:gcd-full} and \eqref{eq:gcd-parity} we find
\begin{equation}
\begin{aligned}
E_{r\,\mathrm{even},\,d\,\mathrm{odd}}(\tau,s)
&=\frac{2^{-s}E(2\tau,s)-2^{-2s}E(\tau,s)}
{1-2^{-2s}},\\[4pt]
E_{r\,\mathrm{odd},\,d\,\mathrm{even}}(\tau,s)
&=\frac{2^{-s}E(\tau/2,s)-2^{-2s}E(\tau,s)}
{1-2^{-2s}},\\[4pt]
E_{r\,\mathrm{odd},\,d\,\mathrm{odd}}(\tau,s)
&=\frac{(1+2^{-2s})E(\tau,s)
-2^{-s}[E(2\tau,s)+E(\tau/2,s)]}{1-2^{-2s}}.
\end{aligned}
\label{eq:parity-reduction}
\end{equation}
At $s=\tfrac12$, the unrestricted $E(\tau,s)$ vanishes at each of
the three moduli $\tau$, $2\tau$, and $\tau/2$. The denominator
is $1/2$, so all three $E_{\mathrm{parity}}(\tau,\tfrac12)$
vanish as well.

To obtain the cutoff sums $F_{\mathrm{parity}}(\tau,\epsilon)$, we insert the identities \eqref{eq:parity-reduction} into the Mellin formula \eqref{eq:mellin-parity} and expand $(1-2^{-2s})^{-1}=\sum_{k\ge0}2^{-2ks}$. The factors of $2^{-s}$ simply rescale the cutoff: for example, $2^{-s}\epsilon^{1/2-s}=2^{-1/2}(2\epsilon)^{1/2-s}$, so $2^{-s}E(\tau,s)$ becomes $2^{-1/2}F(\tau,2\epsilon)$. Applying this rescaling to each term in the geometric series and using absolute convergence for $\Re s>1$ to exchange the sums and integral gives the first two identities below, while the odd-odd class again follows by subtraction: 
\begin{equation}
\begin{aligned}
F_{r\,\mathrm{even},\,d\,\mathrm{odd}}(\tau,\epsilon)
&=\sum_{k=0}^\infty\Big[
2^{-k-1/2}F(2\tau,2\cdot4^k\epsilon)
-2^{-k-1}F(\tau,4\cdot4^k\epsilon)\Big],\\[4pt]
F_{r\,\mathrm{odd},\,d\,\mathrm{even}}(\tau,\epsilon)
&=\sum_{k=0}^\infty\Big[
2^{-k-1/2}F(\tau/2,2\cdot4^k\epsilon)
-2^{-k-1}F(\tau,4\cdot4^k\epsilon)\Big],\\[4pt]
F_{r\,\mathrm{odd},\,d\,\mathrm{odd}}(\tau,\epsilon)
&=F(\tau,\epsilon)-F_{r\,\mathrm{even},\,d\,\mathrm{odd}}(\tau,\epsilon)
-F_{r\,\mathrm{odd},\,d\,\mathrm{even}}(\tau,\epsilon).
\end{aligned}
\label{eq:F-reduction}
\end{equation}
Using Stanford-Yan's result
$F(\tau,\epsilon)=3/\sqrt{\pi\epsilon}+o(1)$, the divergent
terms in each of the first two sums add up to
\[
\frac3{\sqrt{\pi\epsilon}}\sum_{k=0}^\infty
\left(2^{-2k-1}-2^{-2k-2}\right)
=\frac1{\sqrt{\pi\epsilon}}.
\]
For the third class the coefficient is $3-1-1=1$, so the
same subtraction applies to all three.

After subtracting the divergent terms in \eqref{eq:F-reduction}, each unrestricted remainder vanishes as $\epsilon\to0$. The infinite sums also vanish: any finite number of terms tends to zero, while the remaining tails can be made uniformly small because the remainders are bounded and their coefficients decrease as $2^{-k}$. Hence
\[
\lim_{\epsilon\to0}\left[
F_{\mathrm{parity}}(\tau,\epsilon)
-\frac1{\sqrt{\pi\epsilon}}\right]
=E_{\mathrm{parity}}(\tau,\tfrac12)=0
\]
and the subtracted length sums agree with their zeta-regularized values. Matching the full vacuum contribution also requires checking its phase-dependent terms.

\section{Energy-basis check and negative threshold weights}
\label{app:energy}
Appendix \ref{app:parity} determined the cusp divergence by counting primitive filling cycles. Here we provide an independent check in the energy basis, following Stanford-Yan and the spin-sector analysis of BTT \cite{Stanford:2025llj,Boruch:2026hbr}. In \ref{app:cusp-pole}, we first show that the pole of the regulated spectral density at $s=1$ reproduces the cusp subtraction of Section \ref{sec:tau}. In \ref{app:threshold}, we then evaluate the same density at the physical value $s=\frac12$, where its behavior near the spectral threshold produces the finite negative weights. These are different effects: the first is a divergence that affect the entire positive-energy continuum, while the second is a finite contribution that is concentrated at the spectral threshold.

\subsection{Energy-basis check of the cusp subtraction}
\label{app:cusp-pole}

For a nondegenerate Virasoro character, write
\begin{equation}
\chi_P(\tau)=\frac{q^{P^2}}{\eta(\tau)},\qquad
P^2=h-\frac{c-1}{24}.
\label{eq:character}
\end{equation}
The output energy and spin are
\begin{equation}
E=P^2+\bar P^2,\qquad j=P^2-\bar P^2.
\label{eq:energy}
\end{equation}
With the descendant factor $|\eta(\tau)|^{-2}$ removed, the primary
continuum has $E\ge|j|$, with scalar threshold $E=j=0$.

The vacuum character $\chi_{\rm vac}=q^{-(c-1)/24}(1-q)/\eta$
gives four seed terms in its left-right product:
\begin{equation}
\renewcommand{\arraystretch}{1.2}
\begin{array}{c|rrrr}
(h,\bar h)&(0,0)&(0,1)&(1,0)&(1,1)\\ \hline
j'=h-\bar h&0&-1&1&0\\
\text{vacuum sign}&+&-&-&+
\end{array}
\label{eq:four-seeds}
\end{equation}
where $j'=h-\bar h$ labels the seed spin, in contrast to the label $j$ defined in \eqref{eq:energy}, which we reserve for the spin of the output states. The four terms in \eqref{eq:four-seeds} arise from expanding the vacuum factors $(1-q)(1-\bar q)$. The modular kernel rewrites a transformed seed character as an integral over characters in the output energy basis. Writing $P'$ for the seed momentum and $P$ for the output momentum, for $r>0$ we have
\begin{equation}
\begin{split}
\chi_{P'}(\gamma\tau)
&=\int_0^\infty \K^{(\gamma)}_{PP'}\chi_P(\tau)\,dP,\\
\K^{(\gamma)}_{PP'}
&=\varepsilon_\gamma\sqrt{\frac8r}\,
e^{\frac{2\pi\ii}{r}(a(P')^2+dP^2)}
\cos\!\left(\frac{4\pi PP'}r\right).
\label{eq:physical-kernel}
\end{split}
\end{equation}
To regulate the sum over fillings in the energy basis, we introduce the zeta-function regulator $s$ through the following generalization of the modular kernel:
\begin{equation}
\K^{(s,\gamma)}_{PP'}
=2\varepsilon(s,\gamma)\left(\frac{2\pi}{r}\right)^s
\frac{P^{2s-1}}{\Gamma(s)}
e^{\frac{2\pi\ii}{r}(a(P')^2+dP^2)}
{}_0F_1\!\left(s;-\frac{4\pi^2P^2(P')^2}{r^2}\right),
\label{eq:analytic-kernel}
\end{equation}
which reduces to \eqref{eq:physical-kernel} at $s=\tfrac12$. Here $\varepsilon_\gamma$ is the phase arising from the modular transformation of the Dedekind eta function, while $\varepsilon(s,\gamma)$ is its counterpart in the analytically continued kernel, which reduces to $\varepsilon_\gamma$ at $s=\frac12$.
The density is the product of the left- and right-moving kernels, so their modular phases cancel in the absence of a gravitational anomaly.

We are interested in the contribution from large winding numbers, which controls the accumulation divergence. In this limit, at fixed momenta, each hypergeometric factor is $1+O(r^{-2})$. Keeping the constant $1$ from each factor gives the leading density with respect to $dP\,d\bar P$:
\begin{equation}
\left.\rho^{(s)}(P,\bar P)\right|_{\rm leading}
=
\frac{4(2\pi)^{2s}}{\Gamma(s)^2r^{2s}}\,
(P\bar P)^{2s-1}e^{\frac{2\pi\ii}{r}(dj+aj')}.
\label{eq:kernel-leading}
\end{equation}
The corresponding density with respect to $dE\,dj=8P\bar P\,dP\,d\bar P$ is:
\begin{equation}
\left.\rho^{(s)}(E,j)\right|_{\rm leading}
=
\frac{2\pi^{2s}}{\Gamma(s)^2r^{2s}}
(E^2-j^2)^{s-1}e^{\frac{2\pi\ii}{r}(dj+aj')}.
\label{eq:energy-kernel}
\end{equation}
Higher kernel terms contain extra powers of $r^{-2}$ and,
at $j=0$, extra powers of $E$.

We now sum the single-geometry density \eqref{eq:energy-kernel} over the allowed fillings, first without spin restrictions. At fixed $r$, write $d=d_0+nr$, where $0\le d_0<r$, $(d_0,r)=1$, $ad_0=1\pmod r$, and $n\in\Z$. This separates the finite sum over coprime residues from the infinite sum over translations $n$. The phase factors as 
\[
e^{\frac{2\pi\ii}{r}(dj+aj')}
=e^{\frac{2\pi\ii}{r}(d_0j+aj')}e^{2\pi\ii n j}
\]
where only the second factor depends on $n$. Poisson summation gives,
as a distribution in $j$,
\begin{equation}
\sum_{n\in\Z}e^{2\pi\ii n j}
=\sum_{\ell\in\Z}\delta(j-\ell),
\label{eq:bos-poisson}
\end{equation}
so the output spin $j$ is integer. At $j=0$, the remaining finite phase sum is the Ramanujan sum
\begin{equation}
c_r(j')=\sum_{\substack{a\bmod r\\(a,r)=1}}e^{2\pi\ii aj'/r},
\qquad c_r(0)=\varphi(r),\quad c_r(\pm1)=\mu(r),
\label{eq:ramanujan}
\end{equation}
where $\varphi(r)$ counts coprime residues, $\mu(r)$ is the M\"obius function, and we have relabeled the sum over $d_0$ as a sum over $a$. The four vacuum signs give $2\varphi(r)-2\mu(r)$. Using
\begin{equation}
\sum_{r\ge1}\frac{\varphi(r)}{r^{2s}} = \frac{\zeta(2s-1)}{\zeta(2s)},\qquad \sum_{r\ge1}\frac{\mu(r)}{r^{2s}} = \frac1{\zeta(2s)},\qquad \Re s>1,
\label{eq:dirichlet-bos}
\end{equation}
the leading scalar density is
\begin{equation}
\left.\rho_{\rm bos}^{(s)}(E,0)\right|_{\rm leading}
=\frac{2\pi^{2s}}{\Gamma(s)^2}E_+^{2s-2}
\frac{2[\zeta(2s-1)-1]}{\zeta(2s)},
\label{eq:bos-density}
\end{equation}
where $E_+^\alpha=\Theta(E)E^\alpha$ before continuation.

Now impose the spin structure restrictions in \eqref{eq:sectors}. In $R^-$, only odd $r$ contribute and all translations $n$ are allowed, so only the range of the $r$ sum changes. For $NS^+$, $d$ must be odd. If $r$ is even, coprimality makes $d_0$ odd and every translation $n$ is allowed. If $r$ is odd, only every other translation is allowed, with the Poisson summation
\begin{equation}
    \sum_{n\in2\Z+n_0}e^{2\pi\ii n j} = \frac12\sum_{\ell\in\Z} (-1)^{n_0\ell} \delta\left(j-\frac\ell2\right)
\end{equation}
where $n_0$ can be $0$ or $1$.
Thus, in the scalar channel $j=0$, even $r$ contributes with full weight and odd $r$ contributes with half weight. Since the $NS^-$ answer differs by $(-1)^{2j}$, the two NS densities agree at $j=0$. In all cases the finite phase sum in the scalar channel is still $c_r(j')$. Thus, the leading densities at $j=0$ are 
\begin{equation}
\left.\rho_{R^-}^{(s)}(E,0)\right|_{\rm leading}
=\frac{2\pi^{2s}}{\Gamma(s)^2}A_R(s)E_+^{2s-2},\qquad
\left.\rho_{NS^\pm}^{(s)}(E,0)\right|_{\rm leading}
=\frac{2\pi^{2s}}{\Gamma(s)^2}A_{NS}(s)E_+^{2s-2},
\label{eq:sector-densities}
\end{equation}
where
\begin{equation}
\begin{aligned}
A_R(s)&=2\sum_{\substack{r\ge1\\r\ {\rm odd}}}
\frac{\varphi(r)-\mu(r)}{r^{2s}}
=\frac{2\{\zeta(2s-1)(1-2^{1-2s})-1\}}
{\zeta(2s)(1-2^{-2s})},\\[4pt]
A_{NS}(s)&=2\sum_{\substack{r\ge1\\r\ {\rm even}}}
\frac{\varphi(r)-\mu(r)}{r^{2s}}
+\sum_{\substack{r\ge1\\r\ {\rm odd}}}
\frac{\varphi(r)-\mu(r)}{r^{2s}}
=\frac{\zeta(2s-1)-1+2^{1-2s}}
{\zeta(2s)(1-2^{-2s})}.
\end{aligned}
\label{eq:Aclosed}
\end{equation}
These identities hold for $\Re s>1$; the zeta expressions
provide the continuation to the two values of $s$ we need.

We now check that these densities give the cusp subtraction
found by the $E(\tau,s)$ and $F(\tau,\epsilon)$ method in Section~\ref{sec:tau}. The part we need comes from the pole at $s=1$. In
\eqref{eq:Aclosed}, only $\zeta(2s-1)$ diverges there:
\[
\zeta(2s-1)=\frac1{2(s-1)}+O(1),\qquad
\left.\zeta(2s)(1-2^{-2s})\right|_{s=1}=\frac{\pi^2}{8}.
\]
The factor multiplying this zeta function tends to $1$ in
both numerators: $2(1-\tfrac12)=1$ for Ramond, and $1$ for
NS. Dividing by $\pi^2/8$ and including the density prefactor
$2\pi^2$ gives
\begin{equation}
\rho_{R^-}^{(s)}(E,0)=\frac{8\Theta(E)}{s-1}+O(1),\qquad
\rho_{NS^\pm}^{(s)}(E,0)=\frac{8\Theta(E)}{s-1}+O(1).
\label{eq:rho-res}
\end{equation}
For comparison, the bosonic density \eqref{eq:bos-density} gives
\[
\rho_{\rm bos}^{(s)}(E,0)=\frac{12\Theta(E)}{s-1}+O(1),
\qquad s\to1.
\]
Each even spin sector therefore has two thirds of the bosonic
pole coefficient. In each case the divergence is in $s$, not
in $E$: its coefficient is
the same at every $E>0$, rather than concentrated at $E=0$.
The remaining terms in the ${}_0F_1$ expansion of
\eqref{eq:analytic-kernel} and the nonzero-spin densities have
no $s=1$ pole.

To compare with one $E_{\mathrm{parity}}(\tau,s)$ sum in
\eqref{eq:parity-EF}, keep descendants stripped off and
match its regulator by multiplying the energy integral by
$\tau_2^s$:
\begin{equation}
(\tau_2|_\gamma)^s = \frac{\tau_2^s}{|r\tau+d|^{2s}}.
\label{eq:regulator-factor}
\end{equation}
Each sector contains two of the three parity classes; since all three pair sums have the same pole, the classes contribute equally. Dividing by $2$ for the allowed parity classes and by another $2$ for the scalar vacuum terms $1+q\bar q$ isolates the residue of a single $E_{\mathrm{parity}}(\tau,s)$, which has the $s=1$ pole coefficient
\begin{equation}
\frac{\tau_2}{4}
\int_0^\infty 8e^{-2\pi\tau_2E}\,dE
=\frac1\pi.
\label{eq:energy-match}
\end{equation}
This agrees directly with the residue in \eqref{eq:parity-res}.

The same Mellin relation \eqref{eq:mellin-parity} then gives
\begin{equation}
F_{\mathrm{parity}}(\tau,\epsilon)
=\Gamma(\tfrac12)\epsilon^{-1/2}\frac1\pi+o(1)
=\frac1{\sqrt{\pi\epsilon}}+o(1)
\label{eq:energy-cutoff}
\end{equation}
where $\epsilon\to0$, and the vanishing remainder was proved
in Appendix \ref{app:reduction}. Summing the two allowed classes and
restoring $2/(\sqrt{\tau_2}|\eta(\tau)|^2)$ gives the cusp
subtraction in \eqref{eq:main-result}:
\[
Z_{\rm cusp}^{\text{even spin}}(\tau)
=-\frac2{\sqrt{\tau_2}\,|\eta(\tau)|^2}
\frac2{\sqrt{\pi\epsilon}}
=-\frac4{\sqrt{\pi\epsilon\tau_2}\,|\eta(\tau)|^2}.
\]
This matches the cusp subtraction, not the finite threshold
weight computed next.

\subsection{Negative threshold weights}
\label{app:threshold}

We now evaluate the same densities \eqref{eq:sector-densities} at the physical weight $s=\tfrac12$. The arithmetic coefficients in \eqref{eq:Aclosed} vanish there, while the energy dependence develops a pole at $E=0$. Their product leaves a delta-function with finite weight.

To see the endpoint pole, integrate against a smooth,
compactly supported function $f$ on the half-line:
\begin{align}
\int_0^\infty E^{2s-2}f(E)\,dE
&=\frac{f(0)}{2s-1}
+\int_0^1 E^{2s-2}[f(E)-f(0)]\,dE+\int_1^\infty E^{2s-2}f(E)\,dE
\label{eq:endpoint-test}
\end{align}
for $\Re s>\tfrac12$.
Since $f(E)-f(0)=O(E)$, both remainder integrals stay finite
as $s\to\tfrac12$. Thus, as a distribution,
\begin{equation}
E_+^{2s-2}=\frac{\delta(E)}{2s-1}+O_{\rm dist}(1),
\label{eq:endpoint}
\end{equation}
where $\delta(E)$ has unit mass on $[0,\infty)$ and the
remainder stays bounded when integrated against $f$.
This pole comes from the endpoint $E=0$ of the energy integral, rather than from the accumulation of geometries at $s=1$.

From \eqref{eq:Aclosed}, near $s\to\tfrac12$, we have $A_R(s)/(2s-1)\to-4$ and
$A_{NS}(s)/(2s-1)\to-1$. Their zeros cancel the endpoint
pole in \eqref{eq:endpoint}. Including the density prefactor $2\pi^{2s}/\Gamma(s)^2\to2$ gives $-8\delta(E)$ and $-2\delta(E)$ for the full traces. BTT define the Ramond density for a single fermion-parity sector, whereas our $R^-$ partition function includes both sectors. Their density is therefore half of the full Ramond density used here, $\rho_R^{\rm BTT}=\rho_{R^-}/2$, giving
\begin{equation}
\rho_R^{\rm BTT}\big|_{\rm threshold}=-4\delta(E),
\qquad
\rho_{NS^\pm}\big|_{\rm threshold}=-2\delta(E).
\label{eq:BTT-atoms}
\end{equation}
The same conversion would halve the Ramond cusp coefficient if it were expressed per copy.

Only the singular term in \eqref{eq:endpoint} survives as $s\to\frac12$, since the regular remainder vanishes when multiplied by $A_R(s)$ or $A_{NS}(s)$. The higher terms in the ${}_0F_1$ expansion give locally integrable densities and hence no additional threshold delta functions. The resulting finite threshold weights arise from the spin-one vacuum terms $-q-\bar q$ in the Ramond sector and the spin-zero vacuum terms $1+q\bar q$ in the NS sectors.

For comparison, \eqref{eq:bos-density} gives the familiar
$-6\delta(E)$ \cite{Maloney:2007ud,Keller:2014xba}.
The results in a common full-trace normalization are
\begin{equation}
\begin{array}{c|c|c}
 & \text{leading divergence of filling sum}
 & \text{finite threshold weight}\\ \hline
\text{bosonic}&\frac6{\sqrt{\pi\epsilon\tau_2}\,|\eta(\tau)|^2}&-6\delta(E)\\[6pt]
R^-&\frac4{\sqrt{\pi\epsilon\tau_2}\,|\eta(\tau)|^2}&-8\delta(E)\\[6pt]
NS^+,\ NS^-&\frac4{\sqrt{\pi\epsilon\tau_2}\,|\eta(\tau)|^2}&-2\delta(E)
\end{array}
\label{eq:results}
\end{equation}
where the two columns distinguish a divergent sum from a finite
negative weight. For example, $-6\delta(E)$ contributes the
finite quantity $-6|\eta(\tau)|^{-2}$ to the partition function:
the problem is its sign, not convergence. Negative continuous
density above threshold is a further positivity problem, as in
the bosonic odd-spin spectrum \cite{Benjamin:2020mfz}.\footnote{Positivity applies to an ordinary
trace; signs due solely to the $(-1)^{\sf F}$ insertion in
$NS^+$ are not themselves a pathology. We do not analyze
continuum positivity in every spin sector here.} Note that summing over the three even spin sectors gives exactly twice the bosonic result, as expected from gauging fermion parity \cite{Boruch:2026hbr}.

\section{The Eisenstein pole for unequal central charges} \label{app:anomalous-pole}

Here we justify the statement used in Section~\ref{sec:unequal-central-charges}: only vacuum terms with no remaining angular dependence produce the cusp pole. We first show that any residue must be independent of $\tau$, then determine which terms contribute to it.

Fix $\Delta c\in8\Z$. The corrections in \eqref{eq:anomaly-cusp-expansion} contain an extra power of $\tau_2|_\gamma$ and are regular at $s=1$,\footnote{Meromorphic continuation follows by comparison with a standard Selberg-Poincaré series \cite{Keller:2014xba}; exponential damping changes the sum only by terms regular at $s=1$.} so it suffices to consider the Poincaré sum
\begin{equation}
    \mathcal P_J(\tau,s)
    =
    \sum_{\gamma\in\mathbb Z\backslash PSL(2,\Z)}
    e^{\frac{\pi\ii\Delta c}{12}\Phi(\gamma)}
    \left(\tau_2|_\gamma\right)^s
    e^{2\pi\ii(J-\Delta c/24)(\tau_1|_\gamma)}\,,
    \qquad
    J\in\Z\,.
\label{eq:app-anomalous-Poincare}
\end{equation}
The vacuum combination is $2\mathcal P_0-\mathcal P_1-\mathcal P_{-1}$.

Write $R_J(\tau)=\Res_{s=1}\mathcal P_J(\tau,s)$. Why must it be constant? All additional factors are phases, so $|\mathcal P_J(\tau,s)|\le E(\tau,s)$ for real $s>1$: any pole is at most simple and $|R_J(\tau)|\le3/\pi$. To constrain its $\tau$ dependence, apply the hyperbolic Laplacian $\Delta_{\mathbb H}=-\tau_2^2(\partial_{\tau_1}^2+\partial_{\tau_2}^2)$:
\begin{equation}
    \left[\Delta_{\mathbb H}-s(1-s)\right]\mathcal P_J(\tau,s)
    =4\pi^2\left(J-\frac{\Delta c}{24}\right)^2\mathcal P_J(\tau,s+2)\,.
\label{eq:app-laplacian-relation}
\end{equation}
The right-hand side converges absolutely near $s=1$, so taking the residue gives $\Delta_{\mathbb H}R_J(\tau)=0$. The anomaly phase is a finite-order character; on the finite cover where it becomes trivial, $R_J(\tau)$ is a bounded harmonic function and hence constant.\footnote{The cover has finite area. Integrating $R_J^*\Delta_{\mathbb H}R_J=0$ with a cutoff at large cusp height gives $\int|\nabla R_J|^2=0$; boundedness makes the cutoff contribution vanish.} 

Since the residue is constant, it can only be nonzero if the phase acquired from $\tau\mapsto\tau+1$ is trivial. Thus
\begin{equation}
    R_J=e^{-\pi\ii\Delta c/12}R_J
    \qquad\Longrightarrow\qquad
    R_J=0\quad\text{unless }\Delta c\in24\Z\,.
\end{equation}
At these values the anomaly multiplier is trivial, but the angular frequency $J-\Delta c/24$ remains. If it vanishes, $\mathcal P_J(\tau,s)=E(\tau,s)$ and the residue is $3/\pi$.

For a nonzero frequency, we average $\mathcal P_J(\tau,s)$ over $0\le\tau_1<1$, which isolates its zero Fourier mode without changing a constant residue. As in Appendix \ref{app:energy}, we organize the fillings by $r$ and sum over the coprime residues of $d$. Using $\operatorname{Re}(\gamma\tau)=a/r+O_\tau(r^{-2})$, the remaining phase sum becomes the Ramanujan sum $c_r(n)$, where $n=J-\Delta c/24$.\footnote{Using $\gamma\tau=a/r-1/[r(r\tau+d)]$, replace $\tau_1|_\gamma$ by $a/r$. The $O(r^{-2})$ correction gives a series regular near $s=1$. The remaining integral supplies $\sqrt{\pi}\Gamma(s-\tfrac12)\tau_2^{1-s}/\Gamma(s)$, also regular there; the $r=0$ term is entire.} The possible pole is therefore determined by
\begin{equation}
    \sum_{r=1}^\infty\frac{c_r(n)}{r^{2s}}
    =\frac{\sigma_{1-2s}(|n|)}{\zeta(2s)}\,,
    \qquad n\neq0\,.
\label{eq:app-Ramanujan-series}
\end{equation}
Here $c_r$ is defined in \eqref{eq:ramanujan}, and the numerator is the finite divisor sum $\sigma_{1-2s}(|n|)=\sum_{d\mid|n|}d^{1-2s}$, while $\zeta(2s)$ is regular and nonzero at $s=1$. Thus there is no pole at $s=1$, and only the channel with zero angular frequency can contribute to the residue:
\begin{equation}
    \Res_{s=1}\mathcal P_J(\tau,s) = \frac3\pi\,\delta_{\Delta c,24J}\,.
\label{eq:app-selection-rule}
\end{equation}
Restoring the vacuum coefficients $2,-1,-1$ and the descendant factor gives
\begin{equation}
    \Res_{s=1}Z_{\rm bos}^{\mathrm{zeta}}(\tau;s) = \frac{3}{\pi\sqrt{\tau_2}|\eta(\tau)|^2} \left( 2\delta_{\Delta c,0} -\delta_{\Delta c,24} -\delta_{\Delta c,-24} \right)\,.
\end{equation}
Mellin inversion using \eqref{eq:anomaly-mellin} then reproduces the cutoff divergence \eqref{eq:cL-cR-div}.

\subsubsection*{Spin restrictions}
A similar argument applies when a spin structure is fixed. For $NS^+$, the smallest translation preserving the spin structure is $T^2$, so the period is two, and \eqref{eq:spin-anomaly-phase} gives
\begin{equation}
    \Phi_{NS^+}(T^2)
    =2\left[\Phi(T^2)-\Phi(T)\right]=2\,.
\end{equation}
Consequently a nonzero constant residue requires $\Delta c=12m$, $m\in\Z$. This makes the full multiplier trivial, since $\Phi_{\s}$ takes even integer values:
\begin{equation}
    e^{\frac{\pi\ii \Delta c}{12}\Phi_{\s}(\gamma)}
    =e^{\pi\ii m\Phi_{\s}(\gamma)}
    =1\,.
\end{equation}
We must still test the remaining angular frequency. In $\Gamma^0(2)$, $b$ is even, so $ad=1\pmod{2r}$. As above, the sum over coprime residues produces $c_{2r}(2J-\Delta c/12)$, while averaging over $0\le\tau_1<2$ introduces a factor of $1/2$. The possible pole is then determined by
\[
    \frac12\sum_{r\ge1}\frac{c_{2r}(2J-\Delta c/12)}{r^{2s}}\,.
\]
Its argument is integer even for half-integer $J$. For any fixed nonzero integer $n$, $|c_{2r}(n)|\le\sum_{d\mid|n|}d$, so this series converges near $s=1$.

Thus specifying a spin structure weakens the condition for a nonzero residue from $\Delta c\in24\mathbb Z$ to $\Delta c\in12\mathbb Z$, while the remaining angular momentum selection rule still requires $J-\Delta c/24=0$. When $J=\Delta c/24$, the original filling sum reduces to the two allowed parity classes, each with residue $1/\pi$ by \eqref{eq:parity-res}, such that each nonoscillating contribution has residue $2/\pi$. The choices $J=0,\pm1$ give $\Delta c=0,\pm24$ in pure gravity, while $J=0,\pm\tfrac12$ give $\Delta c=0,\pm12$ in supergravity. Restoring the vacuum coefficients and determinants gives \eqref{eq:cL-cR-div-ferm} and \eqref{eq:cL-cR-div-N11}; the other sectors follow by the modular transformations described in Section~\ref{subsec:spin-anomaly}.

\section{Bulk derivation of the spin anomaly phase}
\label{app:spin-anomaly}
Our goal is to derive the phase $e^{\frac{\ii\pi\Delta c}{12}\Phi_{\s}(\gamma)}$ multiplying each filling in \eqref{eq:Z-spin-anom} directly from the bulk gravitational Chern-Simons theory, independently of the boundary calculation in Section~\ref{subsec:spin-anomaly}.\footnote{See also \cite{jeffrey2010, Dabholkar:2014ema, Witten:2019bou, Delmastro:2021xox, Kubo:2021xxi, Witten:2015aba, Boruch:2026hbr} for related references.} We first fix the gluing and framing for $NS^+$ and compute the spin-dependent part of the phase, and then identify $\Phi_{\s}$ for all three sectors. Throughout, $\Delta c\in\tfrac12\Z$.

\paragraph{The bulk phase and its framing.}
We compute the relative gravitational Chern–Simons phase by gluing each filling to an oppositely oriented reference filling, obtaining a closed three-manifold. Start with $NS^+$: the reference filling has a contractible spatial cycle, with periodic boundary conditions in time and antiperiodic boundary conditions in space. For $\gamma=\left(\begin{smallmatrix}a&b\\r&d\end{smallmatrix}\right)\in\Gamma^0(2)$ and $r>0$, use the inverse gluing
\begin{equation}
    U=-\gamma^{-1}=\begin{pmatrix}-d&b\\r&-a\end{pmatrix}
\label{eq:bulk-spin-gluing}
\end{equation}
to compute the bulk phase that cancels the change of the vacuum amplitude under a change of representative. The resulting lens space $Y_\gamma=L(r,-d)$ inherits the spin structure fixed by the boundary conditions.\footnote{Our orientation convention is the quotient of $S^3=\partial B^4$ by $(z_1,z_2)\sim(e^{2\pi\ii/r}z_1,e^{-2\pi\ii d/r}z_2)$. Time and space are $(1/r,-d/r)$ and $(0,1)$ on the gluing torus; the two meridians are therefore space and $r$ times time plus $d$ times space.
We choose the spin lift that follows the rotation of the torus frame along time, so spinor components are periodic. After $r$ repetitions, the two planes make $1$ and $-d$ full turns, giving the spinor sign $(-1)^{1-d}=1$ since $d$ is odd; the lift is therefore consistent. For even $r$, the opposite lift gives the other spin structure, with antiperiodic time rather than the $NS^+$ choice used here \cite{kirby1994,boldt2015}.}

The phase also depends on a framing, which retains the twisting information of the gluing that the lens space alone does not record. We fix the Freed-Gompf surgery two-framing $f$ associated with $U$. Choose an auxiliary spin four-manifold $W$ bounding $Y_\gamma$ and extending its spin structure. Its relative Pontryagin number $\mathcal P_f(W)$ is a curvature integral with a boundary correction fixed by the framing. The bulk theory assigns the phase \cite{Freed:1991wd,Witten:2019bou}
\begin{equation}
    \exp\left[-\frac{\pi\ii\Delta c}{12}\mathcal P_f(W)\right]\,,
    \qquad \mathcal P_f(W)=3\sigma(W)-\Phi(\gamma)\,.
\label{eq:bulk-spin-action}
\end{equation}
Here $\sigma(W)$ is the signature of $W$; the term $-\Phi(\gamma)$ is the framing correction.\footnote{A two-framing trivializes $TY\oplus TY$, with $\mathcal P_f(W)=\tfrac12p_1(2TW,f)$. Atiyah's canonical two-framing gives $\mathcal P_f(W)=3\sigma(W)$ \cite{atiyah1990}; our surgery framing transports the canonical two-framing of $S^3$ through the torus surgeries. Write $-SU=T^{m_k}S\cdots T^{m_1}S$ and let $B$ be the surgery linking matrix. The surgery formula \cite{Freed:1991wd} and the continued-fraction identity \cite[Proposition 3.5]{Jeffrey:1992tk} give $\mathcal P_f(W)-3\sigma(W)=\sum_i m_i-3\sigma(B)=\Phi(U)=-\Phi(\gamma)$.} Changing the auxiliary spin manifold $W$ adds a closed spin four-manifold, for which $\mathcal P=3\sigma\in48\Z$. Since $2\Delta c$ is integer, this leaves the phase unchanged.

With the framing fixed, the signature enters only through its value modulo $16$, the Rokhlin invariant $\mu$ \cite{kirby1994}. The phase therefore becomes
\begin{equation}
    \exp\left[\frac{\ii\pi\Delta c}{12}
        \left(\Phi(\gamma)-3\mu(Y_\gamma)\right)\right]\,,
    \qquad \mu(Y_\gamma)=\sigma(W)\pmod{16}\,,
\label{eq:bulk-spin-rokhlin}
\end{equation}
where the framing term $\Phi(\gamma)$ is known. It remains to compute the spin-dependent quantity $\mu(Y_\gamma)$.\footnote{As consistency check, by taking $\Delta c\in8\Z$ in \eqref{eq:bulk-spin-rokhlin}, we recover precisely the bosonic Rademacher phase in \eqref{eq:Z-bos-anomaly} since $\mu(Y_\gamma)\in\Z$.}

\paragraph{Computing the $NS^+$ phase.}
The Atiyah--Patodi--Singer (APS) index theorem lets us compute $\mu$ from spectral asymmetries on $Y_\gamma$, without constructing $W$ explicitly. For our orientation and spin structure, the Dirac and signature eta invariants are \cite{katase:1987,boldt2015}
\begin{equation}
\begin{aligned}
    \eta_D(Y_\gamma)
    &=\frac1{2r}\sum_{\ell=1}^{r-1}
       \csc\frac{\pi\ell}{r}\,\csc\frac{\pi d\ell}{r}
      =2\left[{\sf s}(d,2r)-{\sf s}(d,r)\right]\,,\\
    \eta_{\rm sign}(Y_\gamma)&=4{\sf s}(d,r)\,,
\end{aligned}
\label{eq:bulk-spin-eta}
\end{equation}
where ${\sf s}$ is the Dedekind sum defined in Section~\ref{subsec:bos}.\footnote{We use the ordinary, not reduced, Dirac eta invariant; the round lens space has no Dirac zero modes. In equivariant APS, the Dirac and signature contributions are $-\tfrac12\csc(\theta_1/2)\csc(\theta_2/2)$ and $-\cot(\theta_1/2)\cot(\theta_2/2)$ \cite{Donelly:1978}. Inserting the quotient angles $2\pi\ell/r$ and $-2\pi d\ell/r$, summing over nonidentity elements, and dividing by $r$ gives the signs above.} With a product metric near the boundary, the APS formulas relate these invariants to the signature \cite{Atiyah:1975jf}:
\begin{equation}
    \operatorname{ind}D_W^+=-\frac1{24}\int_Wp_1-\frac{\eta_D}{2}\,,
    \qquad
    \sigma(W)=\frac13\int_Wp_1-\eta_{\rm sign}\,.
\end{equation}
The spin Dirac index is even, so eliminating $\int_Wp_1$ gives \cite{Miller:1987, Gukov:2025dol}
\begin{equation}
    \mu(Y_\gamma)\equiv-4\eta_D-\eta_{\rm sign}
    =4{\sf s}(d,r)-8{\sf s}(d,2r)\pmod{16}\,.
\end{equation}
Using $\gamma_{NS^+}$ from \eqref{eq:spin-eta-matrices}, we have $\Phi(\gamma)=(a+d)/r-12{\sf s}(d,r)$ and $\Phi(\gamma_{NS^+})=(a+d)/(2r)-12{\sf s}(d,2r)$. The terms involving $a+d$ cancel in $2\Phi(\gamma_{NS^+})-\Phi(\gamma)$. Substituting the resulting relation into \eqref{eq:bulk-spin-rokhlin} gives
\begin{equation}
\begin{aligned}
    3\mu(Y_\gamma)&\equiv2\Phi(\gamma_{NS^+})-\Phi(\gamma)\pmod{48},\\
    e^{\frac{\ii\pi\Delta c}{12}[\Phi(\gamma)-3\mu(Y_\gamma)]}&=
    e^{\frac{\ii\pi\Delta c}{6}[\Phi(\gamma)-\Phi(\gamma_{NS^+})]}.
\end{aligned}
\label{eq:bulk-spin-phase-NS}
\end{equation}
The ambiguity modulo $48$ does not affect the phase because $2\Delta c$ is integer. This completes the $NS^+$ calculation.

\paragraph{Other spin structures and checks.}
The other sectors follow by changing the boundary cycles and transporting the reference filling, spin structure, and framing together. Apply the $NS^+$ calculation to $T\gamma T^{-1}$ for $NS^-$ and to $S\gamma S^{-1}$ for $R^-$; constant phases in the changes of marking cancel. Using the matrices $\gamma_{\s}$ in \eqref{eq:spin-eta-matrices}, the resulting phase is $e^{\frac{\ii\pi\Delta c}{6}[\Phi(\gamma)-\Phi(\gamma_{\s})]}$. Comparing with $e^{\frac{\ii\pi\Delta c}{12}\Phi_{\s}(\gamma)}$ in \eqref{eq:Z-spin-anom} identifies
\begin{equation}
    \Phi_{\s}(\gamma)=2\left[\Phi(\gamma)-\Phi(\gamma_{\s})\right].
\label{eq:bulk-spin-phase-all}
\end{equation}
This reproduces \eqref{eq:spin-anomaly-phase} from the bulk. In particular,
\begin{equation}
    \Phi_{NS^\pm}(T^{2n})=2n,
    \qquad \Phi_{R^-}(T^n)=-2n,
    \qquad \Phi_{R^-}(S^{-1}T^2S)=2\,.
\label{eq:bulk-spin-generator-checks}
\end{equation}
The last relation cancels the $T^2$ phase of the Ramond seed $\chi_{\rm vac}(S\tau)$.

The case $r=0$, excluded from the lens-space calculation, can be checked directly. For $\gamma=T^{2n}$ in $NS^+$, gluing gives $S^2\times S^1$, which bounds $B^3\times S^1$ with $\mu=0$. The modular framing has $\mathcal P_f=-2n$, giving phase $e^{\ii\pi\Delta c\,n/6}$, in agreement with \eqref{eq:bulk-spin-generator-checks}.

\clearpage
\bibliography{ref}
\bibliographystyle{utphys}
\end{document}